\documentclass[smallextended]{svjour3}
\smartqed  % flush right qed marks, e.g. at end of proof
\usepackage{graphicx}

\usepackage{amsmath}
\usepackage{amsfonts}
\usepackage{amssymb}
\usepackage[numbers]{natbib}

\usepackage{stmaryrd}
\usepackage{subfig}

\PassOptionsToPackage{hyphens}{url}\usepackage{hyperref}
\usepackage{graphicx}
\usepackage[ruled,vlined]{algorithm2e}
\usepackage{bm}
\usepackage{xspace}

\DeclareMathOperator{\successor}{succ}
\DeclareMathOperator{\predecessor}{pred}
\DeclareMathOperator{\Win}{Win}

\def\makeheadbox{\relax}

\begin{document}

\newcommand{\stepi}{\textit{i}}
\newcommand{\ii}{\textit{ii}}
\newcommand{\iii}{\textit{iii}}
\newcommand{\iv}{\textit{iv}}
\newcommand{\old}{\textit{old}}

\renewcommand\topfraction{0.85}
\renewcommand\bottomfraction{0.85}
\renewcommand\textfraction{0.20}
\renewcommand\floatpagefraction{0.85}
\setlength{\intextsep}{1\baselineskip}
\newcommand{\hmaster}{\textit{hmaster}}
\newcommand{\hready}{\textit{hready}}
\newcommand{\hbusreqone}{\textit{hbusreq1}}
\newcommand{\hburstzero}{\textit{hburst0}}
\newcommand{\hburstone}{\textit{hburst1}}
\newcommand{\hmasterzero}{\textit{hmaster0}}
\newcommand{\hmastlock}{\textit{hmastlock}}
\newcommand{\hbusreqzero}{\textit{hbusreq0}}
\newcommand{\hgrantzero}{\textit{hgrant0}}
\newcommand{\hgrantone}{\textit{hgrant1}}
\newcommand{\hlockzero}{\textit{hlock0}}
\newcommand{\hlockone}{\textit{hlock1}}
\newcommand{\stateG}{\textit{stateG}}
\newcommand{\stateA}{\textit{stateA}}
\newcommand{\start}{\textit{start}}
\newcommand{\decide}{\textit{decide}}
\newcommand{\locked}{\texts{locked}}
\newcommand{\inx}{\mathcal{X}}
\newcommand{\minx}{$\mathcal{X}$\xspace}
\newcommand{\outy}{\mathcal{Y}}
\newcommand{\mouty}{$\mathcal{Y}$\xspace}
\newcommand{\var}{\mathcal{V}}
\newcommand{\mvar}{$\mathcal{V}$}

\newcommand{\lang}{\mathcal{L}}
\newcommand{\tran}{\mathcal{T}}
\newcommand{\interpolant}{\mathcal{I}}
\newcommand{\game}{\mathcal{G}}
\newcommand{\mgame}{$\mathcal{G}$}
\newcommand{\counterstrategy}{\mathcal{C}}
\newcommand{\unrealcore}{\phi^{uc}}
\newcommand{\envstrat}{\mathcal{E}}

\newcommand{\msys}{$\mathcal{S}$}
\newcommand{\sys}{\mathcal{S}}
\newcommand{\mcont}{$\mathcal{C}$}
\newcommand{\cont}{\mathcal{C}}
\newcommand{\menv}{$\mathcal{E}$}
\newcommand{\env}{\mathcal{E}}
\newcommand{\win}{\textit{Win}}
\newcommand{\fair}{{\scriptsize\textit{fair}}}
\newcommand{\inv}{{\scriptsize\textit{inv}}}
\newcommand{\init}{{\scriptsize\textit{init}}}

% LTL operators and constants
\newcommand{\true}{\textit{true}}
\newcommand{\false}{\textit{false}}

\newcommand{\always}{\textsf{\textbf{G}}}
\newcommand{\eventually}{\textsf{\textbf{F}}}
\providecommand{\next}{\textsf{\textbf{X}}} % \providecommand defines \next only if it has not been defined yet
\newcommand{\until}{\textsf{\textbf{U}}}

% Path states
\newcommand{\initstates}{\states^{init}}
\newcommand{\transstates}{\states^{trans}}
\newcommand{\loopingstates}{\states^{loop}}
\newcommand{\failingstates}{\states^{fail}}
\newcommand{\unrolledstates}{\states^{unr}}
\newcommand{\statesu}{\states_u}

\newcommand{\initstate}{s_0}
\newcommand{\failingstate}{s^{fail}}
\newcommand{\transstate}[1]{s^{trans}_{#1}}
\newcommand{\loopingstate}[1]{s^{loop}_{#1}}
\newcommand{\unrolledstate}[2]{s^{unr}_{#1,#2}}

% Translation
\newcommand{\tcounterplay}{\llbracket\var,\phi^\env\rrbracket_r}
\newcommand{\tcvaluations}{\llbracket\var\rrbracket_r}
\newcommand{\tcassumptions}{\llbracket\phi^\env\rrbracket_r}
\newcommand{\tcguarantees}{\llbracket\phi^{uc}\rrbracket_r}

% Boolean expressions
\newcommand{\boolexpr}[3]{B^{#1}_{#2}(#3)}
\newcommand{\boolxproj}[3]{B^{#1}_{\ifthenelse{\equal{#2}{}}{\inx}{#2,\inx}}(#3)}
\newcommand{\boolyproj}[3]{B^{#1}_{\ifthenelse{\equal{#2}{}}{\outy}{#2,\outy}}(#3)}
\newcommand{\boolexprvar}{\boolexpr{}{}{\var}}
\newcommand{\boolexpression}{B}

% Assumptions sets
\newcommand{\assumptionsinitset}{\Psi_{init}}
\newcommand{\assumptionsinvset}{\Psi_{inv}}
\newcommand{\assumptionsfairset}{\Psi_{fair}}

% Omega-languages expressions
\newcommand{\olanguage}{$\omega$-\-lan\-guage\xspace}
\newcommand{\olanguages}{$\omega$-\-lan\-gua\-ges\xspace}
\newcommand{\oword}{$\omega$-\-word\xspace}
\newcommand{\owords}{$\omega$-\-words\xspace}
\newcommand{\sigmaomega}{\Sigma^\omega}
\newcommand{\sigmastar}{\Sigma^*}
\newcommand{\sigmavar}{\Sigma_\var}

% Buchi automaton symbol
\newcommand{\bautomaton}{\mathcal{B}}
% Muller automaton symbol
\newcommand{\mautomaton}{\mathcal{M}}
% Trachtenbrot automaton symbol
\newcommand{\lautomaton}[1]{\mathcal{A}_{#1}}

\newcommand{\Hausdim}[1]{\dim\left({#1}\right)}
\newcommand{\Hausmeas}[2]{m_{#1}({#2})}
\newcommand{\HausmeasBf}[2]{\mathbf{Haus}_{#1}({#2})}

\newcommand{\preflanguage}[1]{{A({#1})}}
\newcommand{\prefnlanguage}[2]{{A_{#1}({#2})}}
\newcommand{\suff}[2]{\operatorname{S}_{#1}({#2})}
\newcommand{\suffixlanguage}[2]{{\suff{#1}{#2}}}

\newcommand{\entropy}[1]{H({#1})}

% Set-theoretic and topological symbols
%\newcommand{\closure}{\mathcal{C}} %Replace with math operator
\newcommand{\setsize}[1]{{\#({#1})}}

% GR(1) examples
\newcommand{\initial}{\phi^{init}}
\newcommand{\invariant}{\phi^{inv}}
\newcommand{\fairness}{\phi^{fair}}
\newcommand{\cfairness}{\phi^{cfair}} % Fairness complement

% Example variables
\newcommand{\req}{\textit{req}}
\newcommand{\cl}{\textit{cl}}
\newcommand{\gr}{\textit{gr}}
\newcommand{\val}{\textit{val}}

% Set notation
\newcommand{\suchthat}{\; | \;}
\newcommand{\valuations}{\mathcal{I}}
\newcommand{\states}{\mathcal{S}}
\newcommand{\admout}{\valuations_{adm,\outy}}
\title{Interpolation-Based GR(1) Assumptions Refinement}

\author{Davide G. Cavezza \and Dalal Alrajeh}

\institute{Imperial College London, United Kingdom\\
\email{\{d.cavezza15,dalal.alrajeh\}@imperial.ac.uk}
}
\date{}
\maketitle             % typeset the title of the contribution

\begin{abstract}
This paper considers the problem of assumptions refinement in the context of unrealizable
specifications for reactive systems. 
We propose a new counterstrategy-guided synthesis approach for  GR(1) specifications based on Craig's interpolants.
%We show that %Realizability is an important property of specifications for reactive systems. Recent studies 
%have suggested the use of game-theoretic approaches to tackle unrealizability of GR(1) 
%specifications caused by missing environment assumptions. In particular, these consider 
%an iterative use of counterstrategies and application of templates to construct new 
%assumptions that collectively would converge to a realizable specification.
%However these are limited in a number of ways: (i) they are restricted to the class 
%of assumption templates they provide and/or (ii) require users to provide the 
%variables upon which these templates are instantiated. Together these risk 
%resulting in assumptions that do not directly target the cause of unrealizability
% and that are too strong (i.e., over-idealised). Instead we propose in this paper a 
%new counterstrategy-guided approach for synthesizing environment assumptions 
%for GR(1) specifications based on Craig's interpolation. 
Our interpolation-based method   identifies causes for unrealizability and 
computes assumptions that directly target unrealizable cores, 
without the need for user input. Thereby, we discuss how this property reduces the maximum number of steps needed to converge to realizability compared with other techniques. We describe properties of interpolants 
that yield helpful GR(1) assumptions and prove the soundness of the results.
Finally, we demonstrate that our approach yields weaker assumptions than baseline techniques, and finds solutions in case studies that are unsolvable via existing techniques.
\end{abstract}


\section{Introduction}
%The present work describes a novel approach for refining environment specifications in Generalized Reactivity (1). The novelty lies in the use of Craig interpolation to extract those specifications from counterstrategies.

%[Problem area]
%Missing requirements are known to be among the major causes of software failure %in reactive systems 
% \cite{Lamsweerde:2009}. Their incompleteness  
%often results from the conception of over-ideal systems, i.e., where the environment they interact with always behaves as expected \cite{Alrajeh:2012,Lamsweerde:2000}. 
%When adopting this view, unforeseen behaviours may arise once the system is deployed in its environment.
%This view risks  adverse  conditions that may arise once the system is deployed in its environment from being well-accounted for.%, adding to the cost of failure. 
%[Solutions in CEGIS]

Constructing formal specifications of systems that capture user requirements precisely and from which implementations can be successfully derived is a difficult task \cite{Lamsweerde:2009}. Their imprecision  
often results from the conception of over-ideal systems, i.e., where the environment in which the system operates always behaves as expected \cite{Alrajeh:2012,Lamsweerde:2000}. However, in several cases the environment can make one or more requirements impossible to satisfy.
Thus one of the challenges in building  correct specifications is identifying sufficient assumptions over the environment under which a system would always be able to guarantee the satisfaction of its requirements, in other words making a specification \textit{realizable}. 
%This problem has been tackled in  the context of reactive systems. 
%Reactive systems are open systems that interact with the environment and whose successful operation depends on this interaction\cite{Bloem2012}.

%One of the main challenges in the derivation of complete specifications for reactive system is  that  of realizability \cite{Konighofer2009,Cimatti2008}.  
%Reactive systems are open systems that interact with the environment and whose successful operation depends on this interaction\cite{Bloem2012}. 

Automated techniques for generating environment assumptions have been proposed in \cite{Li2011a,Alur2013}. These  make use of counterstrategies to iteratively guide the search for assumptions that would make the specification realizable. (A \emph{counterstrategy} is a characterization of the environment behaviors that force the violation of the specification.)
%
%, which can be computed automatically \cite{Konighofer2009}.
%;in order to rule it out from the set of environments where the specification is expected to be realizable.
%
In each iteration, the specification is first checked for realizability. If it is found to be unrealizable,  a counterstrategy is computed automatically \cite{Konighofer2009}. Then alternative assumption refinements are computed
each of which being inconsistent with the counterstrategy. The alternatives  are again checked for realizability and so forth until a realizable specification is successfully reached or no solutions can be found. 

The problem with existing approaches however is that they heavily rely on the users' knowledge of the problem domain and of the cause of unrealizability.   For instance, the work in \cite{Li2011a} requires users to specify a set of temporal logic templates as formulae with placeholders to be replaced with Boolean variables. Assumptions are then generated as instantiations of such templates that eliminate a given counterstrategy.  This typically constrains the search space to only a class of specifications, which do not necessarily address the cause of unrealizability, and potentially eliminate viable solutions to the realizability problem.
% e.g., by requiring users to provide domain-dependent templates for the assumptions to be computed. 
The work in \cite{Alur2013}, on the other hand, generates such templates automatically. However it requires users to provide a subset of variables to be used for instantiating the templates. This hence puts the burden on the the user to guess the exact
subset of variables that form the cause of unrealizabiliy. This often yields  assumptions that do not target the actual cause of unrealizability, resulting in refinements that needlessly over-constrain the environment.

This paper presents a new counterstrategy-guided inductive synthesis procedure for automatically generating %weak 
assumptions  that instead: (i)  makes use counterstrategies to directly target the cause of unrealizability in a given specification and (ii) does not require users to provide templates or variables' selections  for constructing  assumption refinements. 
We assume an adversarial environment when modelling a system, and focus on specifications expressed in a fragment of  linear temporal logic (LTL) called \emph{Generalized Reactivity (1)} (GR(1) for short).  This subclass is commonly used to express specifications for reactive synthesis \cite{Pnueli1989a}, for which computationally intractable  methods exist  in polynomial time. 

%(The cause of unreliabzibility is defined  in terms of the set of requirements that cannot be guaranteed given the environment assumptions available.)

%Realizability  is concerned with whether  a system  can satisfy a given specification in any environment it may interact with \cite{Pnueli:1989}. Though the problem is known to be 2EXPTIME for general Linear Temporal Logic (LTL)  \cite{Alur2015,Pnueli:1989},
%it has been shown to be tractable for a fragment of LTL, called \emph{Generalized Reactivity (1)} (GR(1)).   A number of algorithms have been proposed 
%for diagnosing realizability  of GR(1) specifications \cite{Konighofer2009,Cimatti2008a}. These consider modelling the realizability problem as a 
%game between two players, the environment and the controller. The controller attempts to satisfy the specification whilst the environment 
%tries to violate it. If the environment can find a winning strategy for the game (referred to as \emph{counterstrategy}) then this is produced as diagnostic feedback. 
%%A specification is said to be unrealizable if no system can implement it.
%%One cause for the unrealizability of a specification may be due to an under-specified environment descriptions. In such case the environment is too permissive, allowing for situations in which the system has no control over its ability to satisfy the specification. 
%Assumptions refinement is concerned with finding environment constraints (\emph{assumptions}) for which there exists an implementation of a controller satisfying the specification.%, making the specification realizable. 

Our procedure iterates over two main phases: realizability check and inductive synthesis.
In brief, it first checks the realizability of a given GR(1) specification comprising assumptions $\phi^\env$ and guarantees $\phi^\sys$. As per existing approaches, if the specification is found to be unrealizable, a counterstrategy is computed. In addition it provides  as output an unrealizable core comprising  a subset of guarantees  from $\phi^\sys$ that are violated in the counterstrategy. 
The key novelty of our procedure is in its use of logical interpolation in the inductive synthesis phase for computing assumptions.   \emph{Craig interpolants}  characterize automatically computable explanations for the inconsistency between Boolean formulae, in their shared alphabet. 
We exploit this feature  to construct expressions  %over variables in the intersection of the assumption and guarantees,  
that \textit{explain} why a counterstrategy, and hence the environment, falsifies a guarantee, and whose negations form  assumptions.
To do so, our procedure translates the  unrealizable core  and the computed counterstrategy  into propositional logic. A Craig interpolant is then computed  to explain the inconsistency between the translated $\phi^\env$ and counterstrategy on one hand and the translated subset of guarantees on the other. The interpolant is then translated  into a GR(1) specification. The resulting formula corresponds to a set of  assumptions that characterize the counterstrategy. Its negation therefore represents an alternative set of assumptions each of which its satisfaction eliminates the counterstrategy. 

To characterize the scope of our approach we introduce the notion of fully-separable interpolants and prove the soundness of our computation when interpolants are fully separable. 
We show that our  proposed  approach is guaranteed to converge to a realizable specification. 
We demonstrate on several case studies that our approach converges more quickly compared to state-of-the-art approaches, namely \cite{Li2011a,Alur2013,Alur2015} by automatically targeting \emph{unrealizable cores} when computing  refinements: that is, the addition of a refinement removes at least one unrealizable core from the original specification in a higher percentage of cases. We further show case studies for which our approaches finds solutions  whilst others fail to do so. Since weakness of assumptions is of importance in reactive synthesis applications, 
we compare the weakness of our refinements with those computed by the  existing techniques.
In summary, our main contributions are:

\begin{itemize}
\item An interpolation-based algorithm for assumption refinement to support reactive synthesis. 
We prove that our proposed procedure terminates in a finite number of steps;
\item We give a definition of fully-separable interpolants which characterizes the class of assumptions produceable through propositional interpolation-based methods;
\item We prove that the assumptions generated in each iteration removes the detected counterstrategy;
\item We show that our procedure finds more solutions than state-of-the-art template-based approaches in a fixed amount of time, despite generating fewer alternatives in each iteration.
\end{itemize}

The rest of this article is organized as follows. Section \ref{sec:Background} introduces relevant background.
Section \ref{sec:refinement} describes the details of the interpolation-based synthesis  approach. Section \ref{sec:Convergence} discusses the convergence of our approach. In Section \ref{sec:Evaluation} we present an evaluation of the proposed method on existing benchmarks, and discuss future directions of improvement in Section~\ref{sec:Discussion}. 
Section \ref{sec:Related} analyzes related work. We conclude the paper in Section \ref{sec:Conclusions}. 

This paper is an extension of our original work \cite{Cavezza2017} in two respects: we present a more comprehensive formalization of the refinement approach, and describe a new experimental setting for a clearer comparison with the state of the art, including new case studies where our approach succeeds where previous ones fail to find any solution.

\section{Background}
\label{sec:Background}
In the following, we use lowercase Latin letters to denote Boolean variables (denoted with the letters $v$, $x$, $y$), infinite sequences (denoted by $w$, $p$), counterstrategy states (denoted by the letter $s$ with subscripts). Uppercase $B$ is used to denote Boolean expressions, while other uppercase letters denote functions. Scripted letters like $\valuations$ and $\var$ denote sets and tuples. Greek letters denote linear temporal logic expressions.

%\noindent
\subsection{Linear Temporal Logic}
Linear temporal logic (LTL) \cite{Manna:1992} is a formalism widely used for specifying reactive systems. The syntax of LTL is defined over a finite non-empty set of propositional variables $\var$, the logical constants \true\ and \false, Boolean connectives, and operators $\next$ (next), $\always$ (always), $\eventually$ (eventually), $\until$ (until). It is described by the BNF expression
$$\phi ::= v | \neg \phi | \phi \land \phi |\phi \lor \phi | \phi \rightarrow \phi | \phi \leftrightarrow \phi | \always \phi | \eventually \phi | \phi \until \phi \:.$$
where $v$ is a terminal symbol belonging to $\var \cup \{\true, \false\}$.

The semantics of LTL consists of infinite sequences of valuations of the variables in $\var$. Such sequences describe formally one observable execution of the system. Let $\valuations = \{I: \var \rightarrow \{\true,\false\}\}$ be the set of all possible valuations of $\var$ (the letter $I$ stands for \emph{interpretation}, as used in \cite{Manna:1992}), and let $\valuations^\omega$ denote the set of infinite sequences of elements from $\valuations$ (this use of the $\omega$ operator comes from the literature on \olanguages, see \cite{Pnueli1989a}). The following rules define when a sequence $w \in \valuations^\omega$ satisfies an LTL formula at position $i \in \mathbb{N}$; $\phi$ and $\psi$ denote any LTL subformula, and $v \in \var$.
\begin{align*}
\langle w, i \rangle & \models \true & \text{always} \\
\langle w, i \rangle & \models \false & \text{never} \\
\langle w, i \rangle & \models v & \text{iff } w_i(v) = \true \\
\langle w, i \rangle & \models \lnot \phi & \text{iff } \langle w, i \rangle \not\models \phi \\
\langle w, i \rangle & \models \phi \lor \psi & \text{iff } \langle w, i \rangle \models \phi \text{ or } \langle w, i \rangle \models \psi \\
\langle w, i \rangle & \models \phi \land \psi & \text{iff } \langle w, i \rangle \models \phi \text{ and } \langle w, i \rangle \models \psi \\
\langle w, i \rangle & \models \next \phi & \text{iff } \langle w, i+1 \rangle \models \phi \\
\langle w, i \rangle & \models \eventually \phi & \text{iff } \exists j \geq i \text{ s. t. } \langle w, j \rangle \models \phi \\
\langle w, i \rangle & \models \always \phi & \text{iff } \forall j \geq i \: \langle w, j \rangle \models \phi \\
\langle w, i \rangle & \models \phi \until \psi & \text{iff } \exists j \geq i \text{ s. t. } \langle w, j \rangle \models \psi \text{ and } \forall i \leq k < j \: \langle w, k \rangle \models \phi
\end{align*} 
For conciseness, we say that $w$ satisfies $\phi$ (in symbols, $w \models \phi$) iff $\langle w, 1 \rangle \models \phi$

In the following we will use a special notation for formulae using Boolean operators only. Given a finite set of terminal symbols $\mathcal{A}$, the expression $B(\mathcal{A})$ denotes a logical formula whose nonterminal symbols are Boolean operators:
$$B(\mathcal{A}) ::= a | \lnot B(\mathcal{A}) | B(\mathcal{A}) \land B(\mathcal{A}) | B(\mathcal{A}) \lor B(\mathcal{A}) | B(\mathcal{A}) \rightarrow B(\mathcal{A}) | B(\mathcal{A}) \leftrightarrow B(\mathcal{A}) $$
where $a \in \mathcal{A}$. We will add superscripts and subscripts to $B$ in order to distinguish different formulae.

The set $\mathcal{A}$ can be a subset of variables or temporal subformulae themselves. Specifically, given $\var' \subseteq \var$, we define $\next \var' = \{\next v \:|\: v \in \var'\}$ the set of terminal symbols obtained by prepending an $\next$ to a variable in $\var'$.

\subsection{Generalized Reactivity (1)}
Generalized reactivity specifications of rank 1 (written GR(1) for short) are a subset of LTL with a specific syntactic structure. Let the variable set $\var$ be partitioned into a set of \emph{input variables} $\inx$ and a set of\emph{ output variables} $\outy$. We define a GR(1) formula as follows.
\begin{definition}
A \emph{generalized reactivity formula of rank 1} (GR(1)) is an LTL formula of the form $\phi^\env \rightarrow \phi^\sys$ The expression $\phi^\env$ is specified as conjunction of one or more of the following subformulae (called \emph{assumptions}):
\begin{enumerate}
	\item a Boolean formula $\varphi^\env_\init$ of the form $B(\inx)$ representing \emph{initial conditions};
	\item a set of LTL formulae $\varphi^\env_\inv$ of the form $\always B(\var\cup \next\inx)$, representing \emph{invariants}; and 
	\item a set of LTL formulae $\varphi^\env_\fair$ of the form $\always\eventually B(\var)$ representing \emph{fairness conditions}.
\end{enumerate}
Likewise, $\phi^\sys$ is a conjunction of the following subformulae (called \emph{guarantees}):
\begin{enumerate}
	\item a Boolean formula $\varphi^\sys_\init$ of the form $B(\var)$ representing \emph{initial conditions};
	\item a set of LTL formulae $\varphi^\sys_\inv$ of the form $\always B(\var\cup \next \var)$, representing \emph{invariants}; and 
	\item a set of LTL formulae $\varphi^\sys_\fair$ of the form $\always\eventually B(\var)$ representing \emph{fairness conditions}.
\end{enumerate}
\end{definition}

We will sometimes indicate GR(1) specifications as a tuple $\langle \phi^\env, \phi^\sys\rangle$ with  $\phi^\theta = \{\varphi^\theta_{\init,i}\} \cup \{\varphi^\theta_{\inv,j}\} \cup \{\varphi^\theta_{\fair,h}\}$ the set of GR(1) units in the formula. Notice that, with a slight abuse of notation, $\phi^\theta$ also denotes the LTL formula obtained by conjoining those units via the $\land$ operator.

Satisfaction of GR(1) formulae by infinite words $w \in \valuations^\omega$ is defined as in general LTL. In the following, we are interested in separating apart the valuation of input and output variables. Given a valuation $I$ and a subset of variables $\var' \subseteq \var$, we denote by $I_{\var'}: \var' \rightarrow \{\true,\false\}$ the restriction of $I$ to $\var'$, that is the valuation defined over $\var'$ such that for every $v \in \var'$ $I(v)=I_{\var'}(v)$. We denote by $\valuations_{\var'}$ the set of all the valuations of variables in $\var'$.

\subsubsection{Co-GR(1) Games}
The formal description of reactive systems is given by two-player game structures. A game structure can be seen as a directed graph such that every state corresponds to some valuation of the system variables and each arc corresponds to a pair of actions available to the two players. The first player, called \emph{environment}, assigns a value to the input variables in order to satisfy the assumptions, and the second player, the \emph{controller}, responds by setting the output variables in compliance with the guarantees. The environment's goal is to force the controller to violate the guarantees while satisfying its assumptions.

In giving the definitions below, we follow the approach by \cite{Konighofer2009}.
\begin{definition}
A \emph{(two-player deterministic) game structure} is a tuple $\game = (\valuations, \Sigma, T, \valuations_0, \Win)$ where
\begin{itemize}
	\item $\valuations = \{I: \var \rightarrow \{\true,\false\}\}$ is the set of game states;
	\item $\Sigma = \valuations_\inx \times \valuations_\outy$ is the set of all possible pairs of input-output valuations;
	\item $T: \valuations \times \Sigma \rightarrow \valuations$ is a transition function, such that for every $I,I' \in \valuations$ $T(I',(I_\inx,I_\outy)) = I$;
	\item $\valuations_0 \subseteq \valuations$ is a set of initial states;
	\item $\Win: \valuations^\omega \rightarrow \{0,1\}$ is a winning condition mapping infinite sequences of states onto a binary value.
\end{itemize}
We call \emph{play} any element $p \in \valuations^\omega$ and say that $p$ is \emph{winning for the controller} if and only if $\Win(p) = 1$. \qed
\end{definition}
With slight abuse of notation, we also call play an element $p \in (\valuations_\inx \times \valuations_\outy)^\omega$ whenever we need to separate the input and output valuations from each other.

Notice we embed states and transitions explicitly as components of $\game$. An alternative may be the symbolic definition provided in \cite{Bloem2012}, where the transitions and the winning condition are replaced by appropriate Boolean formulae.

Automated controller synthesis is achieved by solving a GR(1) game, where the winning condition $\Win$ corresponds to a GR(1) formula $\phi$. Assumptions refinement is needed when such a solution does not exist, in which case a complementary game is of interest, where the winning condition is the negation of the GR(1) formula. In this case the game is called co-GR(1).
\begin{definition}
	A \emph{co-GR(1)} game is a game structure such that:
	$$\forall p \in \valuations^\omega, \; \Win(p) = 1 \text{ if and only if } p \models \phi^\env \land \lnot \phi^\sys$$
\end{definition}

A decision function that, given a finite sequence of game states, returns the environment's next move, is called an \emph{environment strategy}. For co-GR(1) games, the sequence of game states can be replaced by a memory element whose value is to be picked from a finite set $\Gamma$. Following the spirit of \cite{Chatterjee2008}, such strategy can be represented as a \emph{Moore transducer}, an automaton whose transitions are triggered by an input alphabet (which corresponds to $\outy$-valuations in our problem) and whose output is a sequence of symbols from an output alphabet (corresponding to $\inx$-valuations). Formally:
\begin{definition}
	An \emph{environment strategy} is a 4-tuple $\envstrat = (\states,s_0,E,T)$ where
	\begin{itemize}
		\item $\states \subseteq 2^\valuations \times \Gamma$ is a set of states, each corresponding to a set of game states $\valuations_s \subseteq \valuations$ and a memory element $\gamma \in \Gamma$;
		\item $s_0 = (\valuations_0,\gamma_0)$ is the initial state;
		\item $E: \states \rightarrow \valuations_\inx$ is the decision function of the environment, that given a current state returns the next input valuation;
		\item $T: \states \times \valuations_\outy \rightarrow \states$ is the state transition function, that given a current state and an output valuation, returns the next state where the strategy transits.
	\end{itemize}
\end{definition}
An environment strategy determines the set of plays that can be observed over the game graph. We call a \emph{run} of $\env$ a sequence of states $s_0s_1 \dots \in \states^\omega$ such that
\begin{itemize}
	\item $s_0$ is the initial state, and
	\item there exists an output valuation $I_{i,\outy} \in \valuations_\outy$ such that $T(s_{i-1},I_{i,\outy}) = s_i \; \forall i \in \mathbb{N}$
\end{itemize}
We say that a play $p = I_0I_1\dots$ \emph{adheres} to a strategy $\env$ if there is a run $r = s_0s_1\dots \in \states^\omega$ of $\env$ such that $E(s_{i-1}) = I_{i,\inx} \; \forall i \in \mathbb{N}$. We also say that the run $r$ \emph{induces} the play $p$.

We can intuitively define a notion of satisfaction of a GR(1) formula $\phi$ by an environment strategy $\env$. We say that $\env$ satisfies $\phi$ ($\env \models \phi$) if and only if for every play $p$ that adheres to $\env$, $p \models \phi$. We also talk about satisfaction for runs. A run $r = s_0s_1\dots$ satisfies $\phi$ in the state $s_i$ (denoted by $\langle r,s_i\rangle \models \phi$) iff for all plays $p$ induced by $r$, $p$ satisfies $\phi$ in position $i$, that is, $\langle p,i \rangle \models \phi$. 

\subsubsection{Unrealizability, Counterstrategies and Assumptions Refinement}
We define unrealizability as the existence of an environment strategy that wins the co-GR(1) game.
\begin{definition}
	A GR(1) formula $\phi^\env \rightarrow \phi^\sys$ is said to be \emph{unrealizable} if and only if the co-GR(1) game with winning condition $\phi^\env \land \lnot \phi^\sys$ admits an environment strategy that satisfies it. Such strategy is called \emph{counterstrategy}, and is denoted by $\counterstrategy$.
	
	The specification is called \emph{realizable} if it is not unrealizable. \qed
\end{definition}

A counterstrategy is a characterization of environment behaviors that make any controller violate one or more guarantees while satisfying the assumptions.

\begin{example}
	\label{ex:counterstrategyGraph}
	We here define the specification of a very simple request-grant system, which will be a running example throughout the discussion.
	
	Let the input variables be $\inx = \{\req,\cl\}$ and the output variables $\outy = \{\gr,\val\}$. Let $\phi$ be a GR(1) specification with assumption
	$$\fairness_\env = \always\eventually \lnot \req$$
	and guarantees
	$$\invariant_{\sys} = \always(\cl \rightarrow \lnot \val)$$
	$$\fairness_{\sys} = \always\eventually(\gr \land \val) \:.$$
	The specification is unrealizable. The counterstrategy is shown in Figure~\ref{fig:counterstrategyexample1}.
	\begin{figure}
		\centering
		\includegraphics[width=0.7\linewidth]{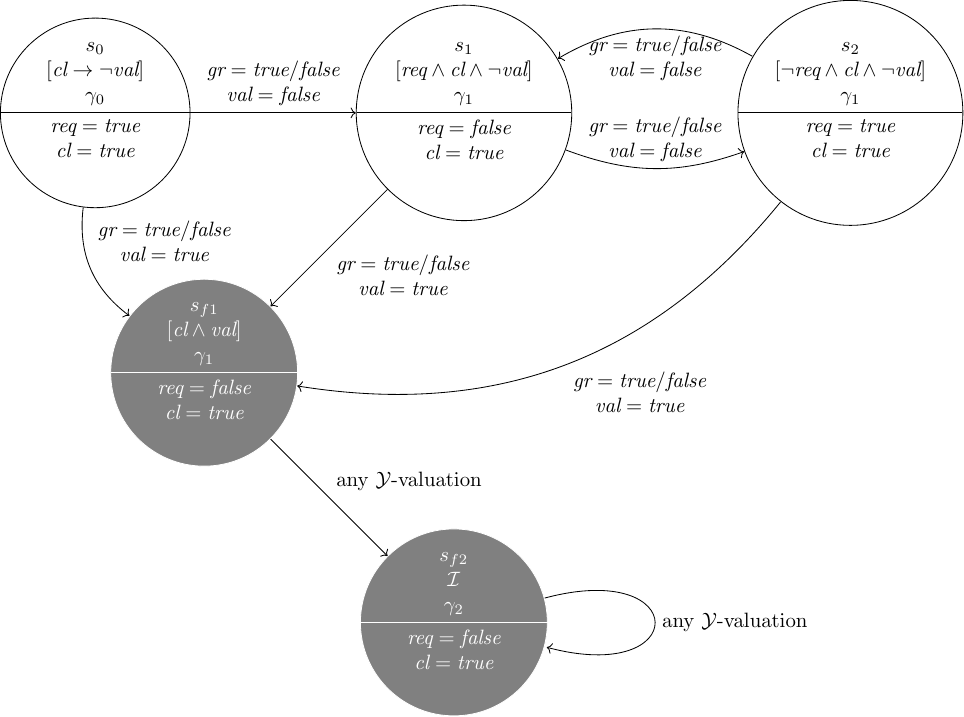}
		\caption{Counterstrategy for specification in Example~\ref{ex:counterstrategyGraph}. The logical expressions in square brackets denote the subset of game states corresponding to each counterstrategy state; $\valuations$ denotes the entire set of possible valuations. The lower half of each state $s$ denotes the next input valuation $E(s)$.}
		\label{fig:counterstrategyexample1}
	\end{figure}
	It represents a set of system behaviors where the environment forces the violation of $\fairness_{\sys}$ by keeping the $\cl$ variable constantly $\true$, while satisfying $\fairness_\env$ by alternating between $\req = \true$ and $\req = \false$. \qed
\end{example}

A GR(1) formula $\phi = \phi^\env \rightarrow \phi^\sys$ is realizable if there are no environment strategies that satisfy $\lnot \phi$. This can be achieved either by weakening $\phi^\sys$, or by strengthening $\phi^\env$. Assumptions refinement deals with the latter option.
\begin{definition}
	Given an unrealizable GR(1) formula $\phi = \phi^\env \rightarrow \phi^\sys$, the problem of \emph{assumptions refinement} requires identifying a set of additional assumptions $\mathbf{\Psi} = \{\psi_1,\dots,\psi_k\}$ such that the refined formula $\phi^\env \land \bigwedge_{i=1}^k \psi_i \rightarrow \phi^\sys$ is realizable.
\end{definition} 
We call \emph{refinements} the additional assumptions $\psi_i$. For simplicity, we call \emph{realizable refinements} those refinements that make $\phi$ realizable, thus solving the realizability problem.

Since a realizable formula is inconsistent with any counterstrategy, and for an unrealizable formula a counterstrategy can be computed automatically \cite{Konighofer2009}, a general approach to compute $\mathbf{\Psi}$ is to generate a set of assumptions that are inconsistent with such a counterstrategies: these approaches are called \emph{counterstrategy-guided assumption refinement} and provide a basis for our work.

\subsubsection{Abstract counterstrategies}
\label{sec:AbstractCounterstrategies}
Typically a full description of a counterstrategy is not needed for assumptions refinement, and more concise descriptions are actually employed \cite{Alur2013,Li2011a,Li2014}. The first simplification consists in removing all controller choices that lead to finite-time violations of the guarantees. The second consists in removing from the transition labellings all the output variables whose assignments are not influential to the choice of the next state (such as $\gr$ in the above example).

Let $\phi_\inv^\sys = \always \boolexpression^\inv(\var \cup \next \var)$ be the invariant guarantee in an unrealizable GR(1) formula $\phi$ (if the formula has more than one invariant guarantee, we can use the equivalence $\always B^\inv_1 \land \always B^\inv_2 \equiv \always(B^\inv_1 \land B^\inv_2)$ to ensure $\phi$ has at most one invariant without loss of generality). Given a state $s = (\valuations_s,\gamma_s) \in \states$ of a counterstrategy, with $\valuations_s = \{I_1, \dots, I_n\}$, let $\valuations_{fail}(s) = \{I: \var \rightarrow \{\true,\false\} \;|\; I_\inx = E(s) \land \forall j \in \{1,\dots,n\}, p = I_jIp_2p_3\dots \; p \not\models \boolexpression^\inv \}$ be the set of all valuations $I$ that cause a violation of the invariant when appearing in a play right after the last valuation $I_j$ observed in $s$. We call this the set of \emph{failing valuations}. Let $\valuations_{fail,\outy}(s) = \{I_\outy: \outy \rightarrow \{\true,\false\} \; | \; I \in \valuations_{fail}(s)\}$ the set of the restrictions of the failing valuations to the output variables. Then, let us call $\admout(s) = \valuations_\outy \backslash \valuations_{fail,\outy}(s)$ the set of all controller responses from $s$ that do not cause a violation of the invariant; we call this the set of \emph{admissible output assignments} from $s$. 

Then we can give the following definition of abstract counterstrategy.
\begin{definition}
	An \emph{abstract counterstrategy} is a 4-tuple $\counterstrategy = (\states,s_0,E,T)$ such that:
	\begin{itemize}
		\item $\states \subseteq 2^\valuations \times \Gamma$ is a set of states;
		\item $s_0 = (\valuations_0,\gamma_0)$ is the initial state;
		\item $E: \states \rightarrow \valuations_\inx$ is the decision function of the environment, that given a current state returns the next input valuation;
		\item $\mathcal{T} = \{T_s\}_{s \in \states}$ is a collection of transition functions, indexed by states in $\states$; for every $s \in \states$, $T_s: \admout(s) \rightarrow \states$ is a function that, given an admissible output assignment for state $s$, returns the next state.
	\end{itemize}
\end{definition}

\begin{example}
	Let us consider again the counterstrategy of Example~\ref{ex:counterstrategyGraph}. A first thing to note is that, since the violation regards a fairness condition, the upper part ends in a strongly connected subset of nodes where $\gr \land \val$ is $\false$.
	
	Notice also that in each state the choice of the next input does not depend on $\gr$, since the state reached in any transition is the same regardless of the value of $\gr$. Moreover, if at any step the controller chooses to set $\val$ to $\true$, the invariant $\invariant_\sys$ is violated: therefore the counterstrategy enters the state $s_{f1}$ and from that point any subsequent infinite play is violating.
	
	Formally,
	$$\valuations_{fail,\outy}(s_0) = \valuations_{fail,\outy}(s_1) = \valuations_{fail,\outy}(s_2) = \{I_\outy \in \valuations_\outy | I_\outy(\val) = \true\}$$
	and
	$$\admout(s_0) = \admout(s_1) = \admout(s_2) = \{I_\outy \in \valuations_\outy | I_\outy(\val) = \false\}$$
	According to the new definition, the counterstrategy will appear as in Figure~\ref{fig:counterstrategyexample2}.
	\begin{figure}
		\centering
		\includegraphics[width=0.7\linewidth]{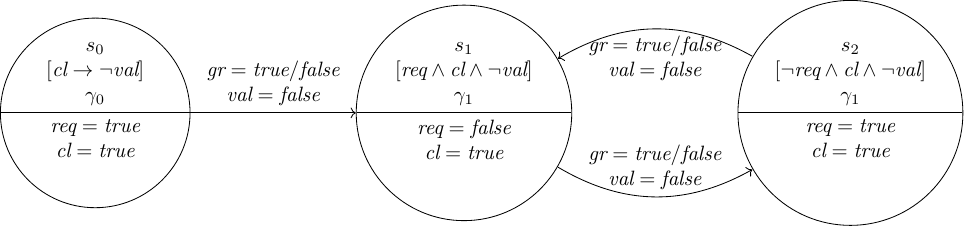}
		\caption{Counterstrategy of the specification in Example~\ref{ex:counterstrategyGraph} with the alternative definition}
		\label{fig:counterstrategyexample2}
	\end{figure}\qed
\end{example}

We finally define the notion of influential output variable for a state.
\begin{definition}
	We define the set of \emph{influential output variables} of a state $s$ as $Y(s) = \{y \in \outy \suchthat \forall I_{1,\outy}, I_{2,\outy} \in \admout(s), \exists I_{1,\outy}, I_{2,\outy} \in \valuations_\outy(s) \text{ such that } I_{1,\outy}(y)\ne I_{2,\outy}(y) \land T_s(I_{1,\outy}) \ne T_s(I_{2,\outy})\}$.
	
	The function $Y: \states \rightarrow 2^\outy$ maps every state to the set of its influential output variables.
\end{definition}
In other words, an influential output variable is a variable whose assignment determines the next state in the counterstrategy. We can provide a more concise description of a counterstrategy by using only influential variables to label transitions.
\begin{example}
	Consider the graph in Figure~\ref{fig:counterstrategyexample2}, and the state $s_0$. We want to identify the set of influential output variables $Y(s_0)$.
	
	First, consider the variable $\val$. As shown in the previous example, $I_\outy \in \admout(s_0)$ if and only if $I_\outy(\val) = \false$. Therefore, there are no two admissible valuations of $\val$ in $\admout(s_0)$, and $\val \not\in Y(s_0)$.
	
	Then consider the variable $\gr$. Both $\true$ and $\false$ appear in some admissible valuations of $s_0$, but the next state does not depend on the value assigned to $\gr$. In symbols:
	$$\forall I_{1,\outy},I_{2,\outy} \in \admout(s_0) \text{ such that } I_{1,\outy}(\gr) \ne I_{2,\outy}(\gr), \; T_{s_0}(I_{1,\outy}) = T_{s_0}(I_{2,\outy}) \;.$$
	Therefore $\gr \not\in Y(s_0)$. We can then conclude that there are no influential output variables in $s_0$, and $Y(s_0) = \varnothing$.
	
	Since the same reasoning can be performed on all other states, the description can be reduced to the simple, one-path graph with unlabeled transitions in Figure~\ref{fig:counterstrategyexample3}.
	\begin{figure}
		\centering
		\includegraphics[width=0.7\linewidth]{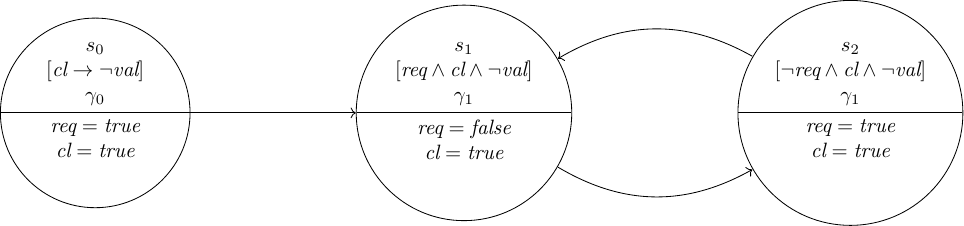}
		\caption{Reduced counterstrategy description for specification in Example~\ref{ex:counterstrategyGraph}}
		\label{fig:counterstrategyexample3}
	\end{figure}
	\qed
\end{example}

Any path on an abstract counterstrategy starting from the initial state is called \emph{counterrun}. It induces a sequence of partial valuations where the subset of variables valuated is different at each step. Given a run $r = s_0s_1s_2\dots$, where $s_0$ is the initial state, we define the \emph{abstract counterplay} (or simply \emph{counterplay}) $p = I_0I_1I_2\dots$ as a sequence of valuations such that
\begin{itemize}
	\item $I_0 \in \valuations_0$
	\item $\forall i > 0, \; I_i: \inx \cup Y(s_{i-1}) \rightarrow \{\true,\false\}$; that is, the $i$-th valuation is defined over the set of input and influential output variables of state $i-1$
	\item $\forall i > 0, \; T_{s_{i-1}}(I_{i}) = s_i$.
\end{itemize}
Counterplays are abstract descriptions of sequences of system states leading to a specific guarantee violation in the counterstrategy. By construction, any play of valuations in $\var$ consistent with an abstract counterplay violates the GR(1) formula $\phi$. Our approach makes use of counterplays to generate assumptions refinements (see Section~\ref{sec:Synthesis}).

\subsubsection{Unrealizable cores}
GR(1) specifications tend to be lengthy and in this case finding the cause of unrealizability is nontrivial. The concept of \emph{unrealizable core} defines a subset of GR(1) conjuncts that help explain the cause of unrealizability. We use a slightly different notion of unrealizable core from the one in \cite{Cimatti2008a}: the original work considered subsetting both assumptions and guarantees, while our definition, following \cite{Konighofer2009}, considers only subsetting guarantees.
\begin{definition}
	Given an unrealizable GR(1) specification $\phi = \langle \phi^\env, \phi^\sys \rangle$, an \emph{unrealizable core} is a subset $\unrealcore \subseteq \phi^\sys$ such that:
	\begin{itemize}
		\item $\langle \phi^\env, \unrealcore \rangle$ is unrealizable, and
		\item for every $\phi' \subseteq \unrealcore$, $\langle \phi^\env, \phi' \rangle$ is realizable.
	\end{itemize}
\end{definition}
For performance reasons, counterstrategy graphs are typically computed from unrealizable cores rather than entire specifications.

%\noindent
\subsection{Craig Interpolation}
\label{sec:Interpolation}
%Craig's interpolation has been used to address partial/incorrect logical theories  in several contexts, such as goal operationalization in KAOS \cite{Degiovanni2014a}, abstraction refinement \cite{Esparza2006}, and model checking \cite{McMillan2003,Vizel2015,DSilva2008}.  
Craig interpolation was originally defined for first-order logic \cite{Craig1957} and later for propositional logic \cite{Krajicek1997}. 
No interpolation theorems have been proved for the general LTL. Extensions have been  proposed recently for LTL fragments  \cite{Kamide2015,Gheerbrant2009}. However these  do not include GR(1) formulae and therefore are not applicable in our case. We use interpolation for propositional logic.

Formally, given an unsatisfiable conjunction of formulae $B_1(\var_1) \wedge B_2(\var_2)$, a Craig interpolant $B_I(\var_I)$ is a formula that is implied by $B_1$, is unsatisfiable in conjunction with $B_2$, and is defined on the common alphabet $\var_I$ of $B_1$ and $B_2$. Recall that a \emph{valid} Boolean formula is a formula that yields $\true$ for any assignment of its variables.

\begin{definition}[Interpolant \cite{Krajicek1997}]
	Let $B_1(\var_1)$ and $B_2(\var_2)$ be two  logical formulae such that their conjunction $\var_1 \wedge \var_2$ is unsatisfiable. Then there exists a third formula $B_I(\var_I)$, called \emph{interpolant} of $B_1$ and $B_2$, such that,  $B_1 \rightarrow B_I$ is valid,   $B_I \rightarrow \lnot B_2$ is valid and $\var_I \subseteq \var_1 \cap \var_2$.
\end{definition}

An interpolant can be considered as an over-approximation of $B_1$ that is still unsatisfiable in conjunction with $B_2$. As stated in Craig's interpolation theorem, although an interpolant always exists, it is not unique.  
Several efficient algorithms have been proposed for interpolation in  propositional logics. 
The resulting interpolant depends on the internal strategies of these algorithms (e.g., SAT solvers, theorem provers). Our approach is based on McMillan's interpolation algorithm  described in \cite{McMillan2003} and implemented in MathSAT  \cite{Cimatti2008}. The algorithm considers a proof by resolution for the unsatisfiability  of $B_1 \wedge B_2$. 

\begin{definition}[Unsatisfiability Proof \cite{McMillan2003}] A proof of unsatisfiability  for a set of clauses $C$ is a directed acyclic graph $(V,E)$, where the vertices $V$ is a set of clauses,  such that for every vertex $c \in V$, either
	\begin{itemize} 
		\item  $c \in C$, and $c$ is a root, or
		\item $c$ has exactly two predecessors, $c_1$ and $c_2$, and  $c$ is their resolvent, 
	\end{itemize} 
	\noindent and
	the empty clause $\false$ is the unique leaf.
\end{definition} 

\begin{definition}[Interpolation Algorithm \cite{McMillan2003}] 
	
	Let $B_1$ and $B_2$ be a pair of clause sets and let $\Pi=(V,E)$ be a proof of unsatisfiability of $B_1 \wedge B_2$ , with leaf vertex \false. For all vertices $c\in V$ , let $p(c)$ be a Boolean formula, such that
	\begin{itemize} 
		\item  if $c$ is a root, then
		\begin{itemize} 
			\item  if $c \in B_1$ then $p(c)=g(c)$,
			\item otherwise  $p(c)$ is the constant \true.
		\end{itemize} 
		\item else, considering $c_1$ and $c_2$ are the predecessors of $c$,  $v$  their pivot variable, 
		\begin{itemize} 
			\item if $v$ is  in the language $\var_1\backslash \var_2$, then $p(c) = p(c_1) \vee p(c_2)$,
			\item otherwise   $p(c) = p(c_1) \wedge p(c_2)$.
		\end{itemize} 
	\end{itemize}
	Then the interpolant is $p(\false)$.
\end{definition} 
%(See Appendix in \cite{DBLP:journals/corr/CavezzaA16} for details.)
%Given two Boolean formulae $A  \vee v  $ and $\neg v  \vee B $, $v$ is called a pivot variable and $A \vee B$ is  their resolvent if it is non-tautological.
  %With this interpretation it has been used in sThe common denominator of these approaches is the use of counterexamples to demonstrate the violation of a property: interpolation is used between the description of a counterexample and the violated property in order to explain what causes the violation in the counterexample. 
%

%As shown in Section~\ref{sec:Counterstrategies}, a counterstrategy must abide by the current assumptions by definition. Therefore, part of its description is given by the assumptions themselves. The other part consists of the valuation of a subset of $\mathcal{V}$ over the traversed states: this subset contains all the input variables and a subset of the output variables that affects the environment's next input choice \cite{Konighofer2009}, and is returned by the counterstrategy computation tool (currently RATSY). This constitutes the term $A$ of interpolation. The term $B$ is given by the guarantees.
%

\section{Approach Overview}
\label{sec:refinement}
Understanding the cause of unrealizability involves checking some execution of the system, that is one particular sequence of interactions between the environment and the controller, identifying the guarantee(s) that was violated and tracing it back to the input assignments that forced that guarantee(s) to be violated. As a concrete instance of this reasoning, let us consider again the specification from Example~\ref{ex:counterstrategyGraph} and the relevant counterstrategy. In that case, the violated guarantee is $\fairness_{\sys}$, since the output variable $\val$ is never $\true$ throughout the infinite path. In turn, the invariant guarantee $\invariant_{\sys}$ expresses a relationship between the input variable $\cl$ and $\val$, that is, whenever $\cl$ is $\true$, $\val$ must be $\false$ by virtue of the implication. By checking the value of $\cl$ along the path, we see $\cl$ is always $\false$ along the path, hence the violation of $\fairness_{\sys}$. An assumption inconsistent with this behavior would be $\always \eventually \cl$, which would make the property realizable.

The first step of the above reasoning, that is identifying a violated guarantee, corresponds to identifying a logical formula to prove; the second step, that is following an implication step to identify a cause of the violation, corresponds to performing a resolution step; finally, the third step of identifying the input variables of interest in the violation corresponds to restraining the cause to the shared alphabet between the assumptions and the guarantees. All these steps are encompassed automatically by Craig interpolation.

\begin{figure}
	\centering
	\includegraphics[width=0.7\linewidth]{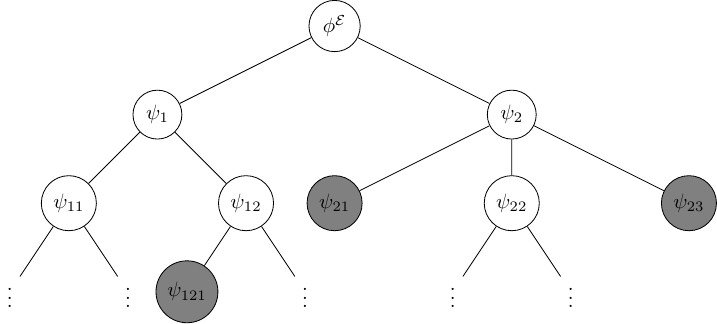}
	\caption{A refinement tree. The refinement corresponding to each node is the conjunction of the $\psi_i$ in the node with all its ancestors up to the root. The grey nodes are realizable leaves.}
	\label{fig:refinementtree}
\end{figure}

The general procedure we propose is based on a sequence of realizability checks and counterstrategy computations,  in the spirit of \cite{Li2011a,Alur2013}, summarized in Algorithm~\ref{alg:Refinements}. A specification $\langle  \phi^\env, \phi^\sys\rangle$ is first checked for realizability. If it is unrealizable, a counterstrategy $\counterstrategy$ and an unrealizable core $\unrealcore$ are computed. The counterstrategy constitutes an example of environment behaviours that force the violation of the guarantees of $\unrealcore$: therefore, the assumptions $\phi^\env$ are refined by adding a GR(1) formula which is inconsistent with the counterstrategy. A set of such formulae $\Psi$ is automatically computed by interpolating ($B_1$) the description of an environment behaviour in the counterstrategy, given by the assumptions and a sequence of state labellings in the counterstrategy;  and ($B_2$) the guarantees.
Each formula $\psi_i \in \Psi$ derived from the interpolant is conjoined in turn to $\phi^\env$ and added to a queue of candidate refinements: then each candidate is checked iteratively for realizability and added to the set of solutions if realizability is achieved; if not, the candidate is further refined and new candidates are added to the queue. In doing this, the procedure produces a tree of candidate refinements, which we will call \emph{refinement tree}, where every node corresponds to a candidate refinement and the leaves correspond to the realizable solutions. The structure of a refinement tree is pictured in Figure~\ref{fig:refinementtree}. By exploring the queue with a first-in-first-out policy, the procedure implements a breadth-first search of the refinement tree.

\begin{figure}[!t]
	\begin{algorithm}[H]
		\caption{\textbf{CounterstrategyGuidedRefinement} procedure}
		\label{alg:Refinements}
		\SetAlgoLined
		\LinesNumbered
		\KwData{$\phi^\env$, assumptions}
		\KwData{$\phi^\sys$, guarantees}
		\KwResult{$refinements = \{\phi^\env \land \psi_i\}$, set of alternative refined assumptions such that $\phi^\env \land \psi_i \rightarrow \phi^\sys$ is realizable for every $i$}
		$candidateRefQueue \gets \{\phi^\env\}$\;
		\Repeat{$candidateRefQueue = \varnothing$}{
			$candidateRef \gets candidateRefQueue$.pop()\;
			\uIf{\textbf{IsSatisfiable}($candidateRef$) and $\lnot$\textbf{IsRealizable}($candidateRef \rightarrow \phi^\sys$)}{
				\label{lalg:RealizabilityCheck}
				$\unrealcore \gets $ \textbf{getUnrealizableCore}$(candidateRef,\phi^\sys)$\;
				$\counterstrategy \gets $ \textbf{getCounterstrategy}$(candidateRef,\phi^\sys)$\;
				$\mathbf{\Psi} \gets $ \textbf{InterpolationBasedSynthesis}$(candidateRef,\unrealcore,\counterstrategy)$\;
				\ForEach{$\psi_i \in \mathbf{\Psi}$}{
					$candidateRefQueue$.append($candidateRef \land \psi_i$)\;
				}
			}
			\uElseIf{\textbf{IsSatisfiable}($candidateRef$)}{
				$refinements$.append($candidateRef$)\;
			}
		}
		\Return $refinements$\;
	\end{algorithm}
\end{figure}

The function \textbf{InterpolationBasedSynthesis} constitutes the core of our proposal (see Algorithm~\ref{alg:InterpolationBasedSynthesis}). It takes as inputs an unrealizable core and a counterstrategy and executes the computation of $\Psi$ via interpolation. 
%The refinement computation is based on Craig interpolation. 
We give the details in the following section.
\begin{figure}[!t]
	\begin{algorithm}[H]
		\caption{\textbf{InterpolationBasedSynthesis}($\phi^\env$, $\unrealcore$, $\counterstrategy$)}
		\label{alg:InterpolationBasedSynthesis}
		\SetAlgoLined
		\LinesNumbered
		\KwData{$\phi^\env$, environment assumptions}
		\KwData{$\unrealcore$, controller guarantees (in an unrealizable core)}
		\KwData{$\counterstrategy$, counterstrategy}
		\KwResult{${\Psi}$, alternative assumptions  eliminating the counterstrategy}
		$r_\counterstrategy$ := \textbf{ExtractCounterrun}($\counterstrategy$)\;
		\label{lalg:ExtractCounterrun}
		$\tcounterplay$:= \textbf{TranslateCounterrunAssumptions}($r_\counterstrategy$, $\phi^\env$)\;
		\label{lalg:TranslateCounterrunAssumptions}
		$\tcguarantees$ := \textbf{TranslateGuarantees}($r_\counterstrategy$,$\varphi^\sys$)\;
		\label{lalg:TranslateGuarantees}
		$I$ := \textbf{Interpolate}($\tcounterplay$, $\tcguarantees$)\;
		\label{lalg:Interpolate}
		\eIf{ $I$ == \false\ or $I$ is not fully-separable}{
			${\Psi}$ := $\{\false\}$\;
		}
		{
			$\mathcal{T}(I)$ := \textbf{TranslateInterpolant}($r_\counterstrategy$, $I$)\;
			\label{lalg:TranslateInterpolant}
			${\Psi}$ := \textbf{ExtractDisjuncts}($\lnot \mathcal{T}(I)$)\;
			\label{lalg:ExtractDisjuncts}
			\Return ${\Psi}$\;
		}
	\end{algorithm}
\end{figure}

\section{Interpolation-Based Synthesis}
\label{sec:Synthesis}
Each execution of \textbf{InterpolationBasedSynthesis} involves extracting temporal formulae that are satisfied by a single run of a counterstrategy (a counterrun), and obtaining refinements from their negation. Excluding a single counterrun is sufficient to make the counterstrategy inconsistent with the refined assumptions. 
Reasoning about counterruns rather than whole counterstrategies has also some advantages, which are discussed in Sect.~\ref{sec:Discussion}.
For the purpose of this paper, we assume that the procedure \textbf{ExtractCounterrun} (line \ref{lalg:ExtractCounterrun}) extracts a counterrun $r_\counterstrategy$ at random.

A counterrun leading to the violation of an initial condition or an invariant guarantee is finite, while that of a fairness guarantee violation ends in a loop \cite{Li2014}. We call the latter a \textit{looping counterrun}, and the loop an \textit{ending loop}.

A counterrun consists of four sets of states: (a)  the \emph{initial state} as the element of the singleton $\initstates = \{\initstate\}$; (b) the  \textit{failing state} in a finite counterrun as the element of $\failingstates = \{\failingstate\}$ 
(c) \textit{looping states} that include the states in ending loop, $\loopingstates = \{\loopingstate{1},\dots,\loopingstate{h}\}$, (d)
\textit{transient states} including all states between the initial state and the first failing state or  loop state (exclusive) $\transstates = \{\transstate{1},\dots,\transstate{k}\}$. %, and finally the  \textit{unrolled states} that capture duplicates of the looping states.
%The set $\unrolledstates$ is initially empty and will be discussed in Sect.~\ref{subsec:Unrolling}. 
With this classification, a finite counterrun has the form $r_\counterstrategy = \initstate\transstate{1}\dots\transstate{k}\failingstate$; whilst a looping counterrun has the form $r_\counterstrategy = \initstate\transstate{1}\dots\transstate{k}(\loopingstate{1}\dots\loopingstate{h})^\omega$. We call $\states_r \subseteq \states$ the union of all the states appearing in $r_\counterstrategy$.

\begin{example}
	Let us consider the specification in Example~\ref{ex:counterstrategyGraph} and the counterstrategy in Figure~\ref{fig:counterstrategyexample3}. The counterstrategy has only one looping run $r_\counterstrategy = s_0(s_1s_2)^\omega$, which is to be used at next stages. This is an example of a looping counterrun with $\initstates = \{s_0\}, \transstates = \varnothing, \loopingstates = \{s_1,s_2\}$, and $\states_r = \states = \{s_0,s_1,s_2\}$. \qed
\end{example} 

%\subsection{Counterplay definition}
%\label{subsec:cpdef}
%By adding to assumptions the negation of a path formula, the environment is prevented from following that path and thereby the whole counterstrategy is ruled out of the game. 

%This operation allows to obtain more specific descriptions of a counterplay in Boolean logic, and therefore allows to compute stronger interpolants. We call \emph{unrolling degree} the number of replicas per state introduced.

%Formally, given a counterstrategy $C=(\Gamma, \gamma_0, \rho)$ in a game \mgame,
%a counterplay $r_\counterstrategy = s_0 s_1 ...$ is the projection of a  counterstrategy play $p = q_0 q_1,...$ over the variables $\inx \cup \outy$  such that  $s_i\in S$ and  $S\subseteq Q$  are the states visited by the run. 

Candidate assumption refinements are computed in four steps: (\stepi) production of two inconsistent Boolean formulae from the counterplay and the unrealizable core, (\ii) interpolation between the two Boolean formulae, (\iii) translation of the interpolant into LTL, and (\iv) negation of the translated interpolant. %and extraction of each disjunct.

\subsection{Boolean Descriptions of Counterplays and Unrealizable Cores}
\label{sec:LTL2Bool}
Step (\stepi) is implemented by the functions \textbf{TranslateCounterrunAssumptions} and \textbf{TranslateGuarantees} (lines \ref{lalg:TranslateCounterrunAssumptions}-\ref{lalg:TranslateGuarantees}). The procedure employs a similar translation scheme as defined in \cite{Biere2003} for bounded model checking, which ensures that the obtained Boolean formula is satisfiable if and only if the play taken into account satisfies the LTL formula. %The inclusion of assumptions in the counterplay translation is important for yielding an interpolant in the shared alphabet of assumptions and guarantees that explains the relationship between the assumptions and the violated guarantees.

The translation of a GR(1) formula into the Boolean domain is a Boolean formula over the domain $\var(\states_r)$ obtained by replicating every variable $v \in \var$ for every state $s \in \states_r$; we denote by $v(s)$ the replica of $v$ referring to state $s$, and by $\var(s)$ the subset of $\var(\states_r)$ containing all the variables referring to state $s$. Formally:
$$\var(\states_r) := \{v(s) | v \in \var, \; s \in \states_r \}$$
$$\forall s \in \states_r, \; \var(s) := \{v(s) | v \in \var\}$$
The translation operation is designed so as to be executed in linear time with the length of $r_\counterstrategy$, and works on GR(1) conjuncts as follows:
\begin{itemize}
	\item an initial condition $\varphi^\theta_{init} = \boolexpr{\init}{}{\var}$ is translated to $\llbracket\varphi^\theta_{init}\rrbracket_r := \boolexpr{\init}{}{\var(\initstate)}$ by replacing each occurrence of $v \in \var$ with $v(\initstate) \in \var(\initstate)$;
	\item an invariant $\varphi^\theta_{inv} = \always \boolexpr{\inv}{}{\var \cup \next \var}$ is translated as the conjunction over all states in $\states_r$ $\llbracket\varphi^\theta_{inv}\rrbracket_r := \bigwedge_{s \in \states_r} \boolexpr{\inv}{}{\var(s) \cup \var(\successor{(s)})}$, where $\successor(s)$ is the successor state of $s$;
	\item a fairness condition $\varphi^\theta_{fair} = \boolexpr{fair}{}{\var}$ is translated into a disjunction over the looping states as $\llbracket\varphi^\theta_{fair}\rrbracket_r := \bigvee_{s \in \loopingstates} \boolexpr{fair}{}{\var(s)}$, or skipped if $r_\counterstrategy$ is not looping.
\end{itemize}

The translation produces two Boolean formulae: a formula $\tcounterplay := \tcassumptions \land \tcvaluations$ that describes $r_\counterstrategy$ and a formula $\tcguarantees$ that describes the unrealizable core. The former consists of two conjuncts:
\begin{itemize}
	\item a Boolean expression $\tcassumptions$ that translates the assumptions by the rules described above;
	\item a Boolean expression $\tcvaluations$ describing the input and output valuations last seen at every state $s$ in $r_\counterstrategy$, restraining the output valuation to the influential output variables for the predecessor of $s$.
\end{itemize}
The expression $\tcvaluations = \land_{s \in \states_r \backslash \initstates} \boolexpression(\var(s))$ is a conjunction of formulae whose variables refer to a single state $s$. Each $\boolexpression(\var(s))$ is a conjunction of literals $v(s)$ or $\lnot v(s)$ for every $v \in \inx \cup Y(\predecessor(s))$, where $\predecessor(s)$ is the predecessor of $s$ in $r_\counterstrategy$; the formula contains the positive literal $v(s)$ if and only if $E(\predecessor(s))(v)=\true$, when $v$ is an input variable, or $I_\outy(v) =\true$ for $T_{\predecessor(s)}(I_\outy) = s$, when $v$ is an influential output variable for $\predecessor(s)$; dually, the formula contains the negative literal $\lnot v(s)$ if and only if $E(\predecessor(s))(v)=\false$, when $v$ is an input variable, or $I_\outy(v) = \false$ for $T_{\predecessor(s)}(I_\outy) = s$, when $v$ is an influential output variable for $\predecessor(s)$. Since by construction the counterplay satisfies the assumptions $\phi^\env$, the formula $\tcounterplay$ is satisfiable.

The formula $\tcguarantees$ only contains the translation of the unrealizable core obtained via the rules described above. Since by definition a counterrun $r_\counterstrategy$ satisfies the assumptions and violates the guarantees, the formula $\tcounterplay \land \tcguarantees$ is unsatisfiable by construction. Therefore, there exists an interpolant for $\tcounterplay$ and $\tcguarantees$.

\begin{example}
	\label{ex:LTL2Boolean}
	In the request-grant protocol, the assumption $\fairness_\env = \always \eventually \lnot \req$ is translated as:
	$$\llbracket\fairness_\env\rrbracket_r = \lnot \req(s_1) \lor \lnot \req(s_2)$$
	Since there is just one assumption, $\tcassumptions = \llbracket\fairness_\env\rrbracket_r = \lnot \req(s_1) \lor \lnot \req(s_2)$
	The formula $\tcvaluations$ is obtained by inspecting Figure~\ref{fig:counterstrategyexample3} for the last valuations of input and influential output variables in each state:
	$$\tcvaluations = (\req(s_1) \land \cl(s_1)) \land (\lnot \req(s_2) \land \cl(s_2)) \;.$$
	The Boolean description of the counterplay is just $\tcounterplay = \tcassumptions \land \tcvaluations$.
	
	The translation of the invariant guarantee $\invariant_\sys = \always(\cl \rightarrow \lnot \val)$ is:
	$$\llbracket\invariant_\sys\rrbracket_r = (\cl(s_0) \rightarrow \lnot \val(s_0)) \land (\cl(s_1) \rightarrow \lnot \val(s_1)) \land (\cl(s_2) \rightarrow \lnot \val(s_2)) \;.$$
	That of the fairness guarantee $\fairness_{\sys} = \always\eventually(\gr \land \val)$:
	$$\llbracket\fairness_\sys\rrbracket_r = (\gr(s_1) \land \val(s_1)) \lor (\gr(s_2) \lor \val(s_2)) \;.$$
	Hence, the translation of the guarantees is just
	$$\tcguarantees = \llbracket\invariant_\sys\rrbracket_r \land \llbracket\fairness_\sys\rrbracket_r$$ \qed
\end{example}

\subsection{Interpolation and Full Separability}

Step (\ii) consists of the function \textbf{Interpolate} (line \ref{lalg:Interpolate}). The returned interpolant $I$ is an over-approximation of $\tcounterplay$ which by definition implies the negation of $\tcguarantees$: it can be interpreted as a cause of the guarantees not being satisfied by the counterplay, and as such a characterization of a set of counterplays not satisfying the guarantees.

From such interpolant the procedure aims at extracting a set of refinements that fit the GR(1) format. In order to do this, the Boolean to temporal translation requires the interpolant to adhere a specific structure. This is embodied in the notion of \textit{full-separability}. To formally define full-separability, we need first to define state-separability and I/O-separability.

\begin{definition}[State-separable interpolant]
	An interpolant $I$ is said to be \emph{state-separable} iff it is in the form
	\begin{equation}
	\bigwedge_{s \in \states_r} \boolexpr{}{s}{\var(s)}
	\end{equation}
	where $\boolexpr{}{s}{\var(s)}$ is a Boolean formula either equal to $\true$ or expressed over variables in $\mathcal{V}(s)$ only.
\end{definition}
We will refer to each $\boolexpr{}{s}{\var(s)}$ as a \emph{state component} of the interpolant. In particular, a state component is equal to $\true$ if $I$ does not use any variables from $s$. State-separability intuitively means that the subformulae of the interpolant involving every single state in the counterplay are linked by conjunctions. This means that in any model of the interpolant each state component must be itself \true.

\begin{definition}[I/O-separable Boolean expression]
	A Boolean expression $\boolexpr{}{s}{\var(s)}$ is said to be \emph{I/O-separable} if it can be written as a conjunction of two subformulae containing only input and output variables respectively:
	\begin{equation}
	\boolexpr{}{s}{\var(s)} = \boolxproj{}{s}{\inx(s)} \land \boolyproj{}{s}{\outy(s)}
	\end{equation}
\end{definition}
We call $\boolxproj{}{s}{\inx(s)}$ and $\boolyproj{}{s}{\outy(s)}$ the \emph{projections} of $\boolexpr{}{s}{\var(s)}$ onto \minx and \mouty respectively. Any model of an I/O-separable Boolean expression satisfies the projections separately.
We can now define full-separability of an interpolant.
\begin{definition}[Fully-separable interpolant]
	An interpolant is called \emph{fully-separable} if it is state-separable and each of its state components is I/O-separable.
\end{definition}

An example of a fully-separable interpolant over $\inx = \{a,b\}, \outy = \{c,d\}$ and states $S=\{s_0,s_1\}$ is $(a(s_0) \lor b(s_0)) \land c(s_0) \land \lnot b(s_1)$; a non-fully-separable interpolant, instead, is $a(s_0) \lor a(s_1)$, since literals referring to different states are linked via a disjunction.

\begin{remark}
	 A particular class of fully-separable interpolants is that of fully conjunctive interpolants, where no disjunctions appear. Whether or not the resulting interpolant is conjunctive depends on the order in which the interpolation algorithm \cite{McMillan2003} chooses the root clauses for building the unsatisfiability proof.
	A sufficient condition for obtaining a fully-conjunctive interpolant is that such root clauses be single literals from $\tcounterplay$, and that the pivot variable in each resolution step belong to the shared alphabet of $\tcounterplay$ and $\tcguarantees$ (see Section~\ref{sec:Interpolation}).
\end{remark}

\begin{example}
	The interpolant of $\tcounterplay$ and $\tcguarantees$ from Example~\ref{ex:LTL2Boolean} is $I = \cl(s_1) \land \cl(s_2)$, which is fully separable. Notice that $I$ captures the environment's choices that cause the violation: $\cl$ being always $\true$ in the looping states $s_1$ and $s_2$. \qed
\end{example}

\subsection{Interpolant Translation}
Step (\iii) consists of the function \textbf{TranslateInterpolant} (line~\ref{lalg:TranslateInterpolant}). It converts a fully-separable interpolant
%\begin{equation}
%\label{eq:Interpolant}
	$I = \bigwedge_{s \in \states_r} \boolexpr{}{s}{\var(s)}$
%\end{equation}
into the LTL formula
\begin{equation}
\label{eq:InterpolantTranslation}
	\begin{split}
	\mathcal{T}(I) := & \boolxproj{init}{}{\inx} \land 
	\bigwedge_{s \in \states_r} \eventually \left( \boolexpr{}{s}{\var} \land \boolxproj{}{\successor{(s)}}{\next \inx} \right) \land \\
	&\eventually \always \bigvee_{j=1}^{|\loopingstates|} \boolexpr{loop}{j}{\var} %\land \bigwedge_{r=1}^u \boolexpr{unr}{j,r}{\var} \right)
	\end{split}
\end{equation}
where the expression $\boolxproj{init}{}{\inx}$ is a shorthand for $\boolxproj{}{\initstate}{\inx}$ and $\boolexpr{loop}{j}{\var}$ for $\boolexpr{}{\loopingstate{j}}{\var}$. %and $\boolexpr{unr}{j,r}{\var}$ for $\boolexpr{}{\unrolledstate{j}{r}}{\var}$. 
Formula (\ref{eq:InterpolantTranslation}) is formed from the single state components of $I$ by replacing the variables in $\var(s)$ with the corresponding variables in $\var$ and by projecting the components onto the input variables where required by the GR(1) template. The translation consists of three units: a subformula describing the initial state, a conjunction of $\eventually$ formulae each containing two consecutive state components, and an $\eventually\always$ formula derived from the looping state components of $I$.

Formula (\ref{eq:InterpolantTranslation}) is guaranteed to hold in the counterplay $r_\counterstrategy$. Intuitively, since $I$ is fully-separable  by construction, $\tcounterplay$ implies each state component and its projections onto \minx and $\outy$. A state component $\boolexpr{}{s}{\var(s)}$ corresponds to a formula $\boolexpr{}{s}{\var}$ satisfied by state $s$ of the counterplay. Therefore, since the initial state satisfies $\boolexpr{init}{}{\var}$, $r_\counterstrategy$ satisfies $\boolxproj{init}{}{\inx}$; since there are two consecutive states $s$ and $\successor{(s)}$ that satisfy $\boolexpr{}{s}{\var(s)}$ and $\boolexpr{}{s}{\var(\successor{(s)})}$ respectively, $r_\counterstrategy$ satisfies $\eventually \left( \boolexpr{}{s}{\var} \land \boolxproj{}{\successor{(s)}}{\next \inx} \right)$. Finally, for the $\eventually\always$ subformula, it is sufficient to observe that the looping state $j$ satisfies the formula $\boolexpr{loop}{j}{\var}$: since the counterplay remains indefinitely in each of the looping states, there is a suffix of it where such formula is true for at least one $j$. Based on these considerations, we  prove the following soundness property. %\footnote{full-separability is a required condition of the interpolant in order to prove the soundness of Formula (\ref{eq:InterpolantTranslation}).}
\begin{theorem}
\label{th:soundness}
	Let $r_\counterstrategy$ be a counterplay and $\phi^\env$ a set of assumptions satisfied in $r_\counterstrategy$, such that their Boolean translation $\tcounterplay$ implies $I$, and let $I$ be a fully-separable interpolant. Then $r_\counterstrategy \models \mathcal{T}(I)$.
\end{theorem}
%\begin{proof}
%See Appendix.
%\end{proof}

\proof Since we are assuming that $I$ is state-separable, and since by definition $\tcounterplay$ implies $I$, each state component (which is a conjunct in $I$) is implied by $\tcounterplay$:
\begin{equation}
\tcounterplay \rightarrow I \rightarrow \boolexpr{}{s}{\var(s)}
\end{equation}
for every $s$.
By construction, a state component holds true iff it is satisfied by the corresponding state in the counterrun:
\begin{equation}
\label{eq:BooleanSatisfaction}
\tcounterplay \rightarrow \boolexpr{}{s}{\var(s)} \Leftrightarrow \langle r_\counterstrategy, s \rangle \models \boolexpr{}{s}{\mathcal{V}}
\end{equation}

Now let us consider $\initstate$. By (\ref{eq:BooleanSatisfaction}) and I/O-separability:
\begin{equation}
\begin{split}
\tcounterplay \rightarrow \boolexpr{init}{}{V(\initstate)} \Rightarrow & \langle r_\counterstrategy, \initstate \rangle \models \boolexpr{init}{}{V} \\
\Rightarrow & \langle r_\counterstrategy,\initstate \rangle \models B^{init}_\inx(\mathcal{X}) \land B^{init}_\outy(\mathcal{Y}) \\
\Rightarrow & r_\counterstrategy \models B^{init}_\inx(\mathcal{X})
\end{split}
\end{equation}
So, $r_\counterstrategy$ satisfies the part of the translation that refers to the initial state.

The next step is to consider pairs of consecutive states $s$ and $\successor{(s)}$. %We have:
%	\begin{equation}
%	\label{eq:currentState}
%		B_s(\mathcal{V}(s)) \Rightarrow  \langle r_\counterstrategy, s \rangle \models B_s(\mathcal{V})
%	\end{equation}
Since $I_u$ is I/O-separable, we have
\begin{equation*}
B_{\successor (s)}(\mathcal{V}(\successor(s))) \rightarrow B_{\successor(s),\inx}(\mathcal{X}(\successor(s))) 
\end{equation*}

By (\ref{eq:BooleanSatisfaction}) and LTL satisfaction definition for $\mathbf{X}$:
\begin{equation}
\label{eq:nextState}
\begin{split}
\tcounterplay \rightarrow B_{\successor(s),\inx}(\mathcal{X}(\successor(s))) \Rightarrow & \langle r_\counterstrategy, \successor(s) \rangle \models B_{\successor(s),\inx}(\mathcal{X}) \\
\Rightarrow & \langle r_\counterstrategy, s \rangle \models \mathbf{X} B_{\successor(s),\inx}(\mathcal{X}) \\
\Rightarrow & \langle r_\counterstrategy, s \rangle \models B_{\successor(s),\inx}(\mathbf{X}\mathcal{X})
\end{split}
\end{equation}

From the conjunction of (\ref{eq:BooleanSatisfaction}) and (\ref{eq:nextState}), and from the LTL interpretation of the operator $\mathbf{F}$ we finally get
\begin{equation}
\begin{split}
\tcounterplay \rightarrow & B_s(\mathcal{V}(s)) \land B_{\successor (s)}(\mathcal{V}(\successor(s))) \\ \Rightarrow &\langle r_\counterstrategy, s \rangle \models B_s(\mathcal{V}) \land B_{\successor(s),\inx}(\mathbf{X}\mathcal{X}) \\
\Rightarrow & r_\counterstrategy \models \mathbf{F} \left( B_s(\mathcal{V}) \land B_{\successor{(s)},\inx}(\mathbf{X}\mathcal{X}) \right)
\end{split}
\end{equation}
Therefore $r_\counterstrategy$ satisfies the ``eventually'' subformulae in the translation.

Finally, let us consider the looping and unrolled states. In general, a path ending with a loop among states $s_1, \dots, s_{|\loopingstates|}$ satisfies the formula
\begin{equation}
\label{eq:templateLoopSatisfaction}
\mathbf{FG} \bigvee_{j=1}^{|\loopingstates|} B_{j}(\mathcal{V})
\end{equation}
where $B_{j}$ is any Boolean expression that holds in $s_j$. The reason is that there is a suffix of the path that contains only states from $\loopingstates$; therefore, this suffix always satisfies any of the looping states' valuations. 

%Now, by construction, unrolled states are replicates of looping states; therefore any Boolean formula that is true in an unrolled state is also true in the looping state of which it is a replicate:
%\begin{equation}
%\langle r_\counterstrategy,s^{loop}_j \rangle \models B^{unr}_{j,r}(\mathcal{V}) \Leftrightarrow \langle r_\counterstrategy,s^{unr}_{j,r} \rangle \models B^{unr}_{j,r}(\mathcal{V})
%\end{equation}

Hence, given
\begin{equation}
\label{eq:satisfiedInLoopingState}
\langle r_\counterstrategy,s^{loop}_j \rangle \models B^{loop}_j(\mathcal{V})
\end{equation}
this can be replaced to the formula in (\ref{eq:templateLoopSatisfaction}) and we obtain:
\begin{equation}
r_\counterstrategy \models \mathbf{F} \mathbf{G} \bigvee_{j=1}^{|S^{loop}|} B^{loop}_j(\mathcal{V})
\end{equation}
%Therefore $r_\counterstrategy$ satisfies also the conjuncts of $\mathcal{T}(I_u)$ that refer to looping states.
%	From this and (\ref{eq:BooleanSatisfaction}) we conclude that
%	\begin{equation}
%	\label{eq:LoopTranslation}
%	r_\counterstrategy \models \mathbf{FG} \bigvee_{j=1}^{|S_{loop}|} B_{loop_j}(\mathcal{V})
%	\end{equation}
%	in case no unrolling is performed (that is there is a bijective mapping between $q_j$ and $s_{loop_j}$). If unrolling takes place, there is at least a state $q_{j'}$ that is mapped onto more than one state in the Boolean domain, let these states be $\{s_loop_{j'},\dots, s_loop_{j'+l}\}$.
%	\begin{equation}
%	\bigwedge_{h=0}^l B_{loop_{j'+h}}(\mathcal{V}(s_{j'+h}))
%	\end{equation}
%	From (\ref{eq:BooleanSatisfaction}), back in the temporal domain we obtain
%	\begin{equation}
%	\langle r_\counterstrategy, q_{j'} \rangle \models \bigwedge_{h=0}^l B_{loop_{j'+h}}(\mathcal{V})
%	\end{equation}
%	But this implies
%	\begin{equation}
%	\label{eq:UnrolledStateBooleanExpression}
%	\langle r_\counterstrategy, q_{j'} \rangle \models \bigvee_{h=0}^l B_{loop_{j'+h}}(\mathcal{V}) = B_{q_{j'}}(\mathcal{V})
%	\end{equation}
%	Therefore, even in case of an unrolling $r_\counterstrategy$ satisfies an $\mathbf{FG}$ formula that can be written as a disjunction of all the $B_{loop_j}$ that appear in the interpolant, as in (\ref{eq:LoopTranslation}).

Since $r_\counterstrategy$ satisfies each of the conjuncts in (\ref{eq:InterpolantTranslation}), then $r_\counterstrategy \models \mathcal{T}(I)$. \qed
\endproof
 
In the case a fully-separable interpolant is not generated from which  $\mathcal{T}(I)$  can be constructed, the algorithm returns $\false$ as its candidate assumption. Otherwise, the approach proceeds 
 to step (\iv) (function \textbf{ExtractDisjuncts}, line~\ref{lalg:ExtractDisjuncts}) producing the candidate refinements by negating (\ref{eq:InterpolantTranslation}) and extracting the disjuncts in the resulting formula:
\begin{equation}
\label{eq:Refinements}
\begin{split}
\lnot \boolxproj{init}{}{\inx} \lor 
\bigvee_{s \in S_u} \always \lnot \left( \boolexpr{}{s}{\var} \land \boolxproj{}{\successor{(s)}}{\next \inx} \right) \lor \\
\always \eventually \bigwedge_{j=1}^{|\loopingstates|} \lnot \boolexpr{loop}{j}{\var} %\left(\boolexpr{loop}{j}{\var} \land \bigwedge_{r=1}^u \boolexpr{unr}{j,r}{\var} \right)
\end{split}
\end{equation}

Each disjunct above is a GR(1) candidate assumption which, by Theorem~\ref{th:soundness},  ensures the exclusion of the counterplay $r_\counterstrategy$ from the models of the assumptions. 

\subsection{Unrolling}
\label{subsec:Unrolling}

\begin{figure}[!t]
	\begin{algorithm}[H]
		\caption{\textbf{InterpolationBasedSynthesis}($\varphi^\env$, $\varphi^\sys$, $C$) with unrolling}
		\label{alg:InterpolationBasedSynthesisWithUnrolling}
		\SetAlgoLined
		\LinesNumbered
		\KwData{$\phi^\env$, environment assumptions}
		\KwData{$\unrealcore$, controller guarantees (in an unrealizable core)}
		\KwData{$\counterstrategy$, counterstrategy}
		\KwResult{${\Psi}$, alternative assumptions  eliminating the counterstrategy}
		$r_\counterstrategy$ := \textbf{ExtractCounterrun}($\counterstrategy$)\;
		\label{lalg:unrExtractCounterrun}
		$u$ := 0\;
		\label{lalg:unrUInit}
		$r_{\counterstrategy,u}$ := $r_\counterstrategy$\;
		\label{lalg:zerounroll}
		${\Psi_{old}}$ := $\emptyset$\;
		%\textit{stopping\_condition} := \true\;
		\Repeat{stopping\_condition}{
			$\tcounterplay$:= \textbf{TranslateCounterrunAssumptions}($r_{\counterstrategy,u}$, $\phi^\env$)\;
			\label{lalg:unrTranslateCounterrunAssumptions}
			$\tcguarantees$ := \textbf{TranslateGuarantees}($r_{\counterstrategy,u}$,$\unrealcore$)\;
			\label{lalg:unrTranslateGuarantees}
			$I_u$ := \textbf{Interpolate}($\tcounterplay$, $\tcguarantees$)\;
			\label{lalg:unrInterpolate}
			\eIf{ $I_u = \false$\ or $I_u$ is not fully-separable}{
				${\Psi}$ := $\{\false\}$\;
				\textit{stopping\_condition} := \textbf{CheckStoppingCondition}($\Psi,\Psi_{old},u$)\;
			}
			{
				$\mathcal{T}(I_u)$ := \textbf{TranslateUnrolledInterpolant}($r_{\counterstrategy,u}$, $I_u$)\;
				\label{lalg:unrTranslateInterpolant}
				${\Psi}$ := \textbf{ExtractDisjuncts}($\lnot \mathcal{T}(I_u)$)\;
				\label{lalg:unrExtractDisjuncts}
				\eIf{$r_{\counterstrategy,u}$ is looping}{
					${\Psi_{old}}$ := ${\Psi}$\;
					$u$ := $u+1$\;
					$r_{\counterstrategy,u}$ := \textbf{UnrollCounterrun}($r_\counterstrategy$,$u$)\;
					\label{lalg:UnrollCounterrun}
					$stopping\_condition$ := \textbf{CheckStoppingCondition}($\Psi,\Psi_{old},u$)\;
				}
				{
					\textit{stopping\_condition} := $\true$\;
				}
				\label{lalg:EndEquivalenceCheck}
				
			}
		}\label{lalg:stoppingCondition} 
		\Return ${\Psi}$\;
	\end{algorithm}
\end{figure}

%In the case of a finite counterplay, the inner-cycle of the synthesis step terminates and outputs the candidate refinements computed above. 
%The equivalence checking of the produced candidates and the unrolling of the counterplay (lines \ref{lalg:EquivalenceCheck}-\ref{lalg:EndEquivalenceCheck}) are only executed in case of a looping counterplay. Thus in each iteration of the inner-cycle, our procedure checks whether the synthesized assumptions are equivalent to the assumptions  $\boldsymbol{\psi_\old}$ computed in the previous iteration. 
%
%If not, the looping part of the counterplay is unrolled once (\textbf{UnrollCounterplay}, line~\ref{lalg:UnrollCounterplay}) and the steps in Sect.~\ref{subsec:refcomp}--\ref{subsec:Unrolling} are repeated. If the equivalence condition is met, the synthesis procedure returns the last set of computed candidates as output.
A common problem to assumptions refinement approaches is that of the \emph{vacuity} of the refined assumptions \cite{Beer2001}. A GR(1) specification is said to be \emph{vacuously realizable} if its assumptions are unsatisfiable; in this case any controller satisfies the specification, since the assumptions evaluate to $\false$. In some cases assumptions refinement approaches trivially eliminate counterstrategies by adding a refinement $\psi$ that is inconsistent with previous assumptions, making the specification vacuously realizable.

Interpolation-based synthesis produces a number of assumptions that grows proportionally with the number of state components in the interpolant, and thereby with the number of states in the counterrun. If the counterrun is small, few refinements are produced and all of them may be inconsistent with the original assumptions. In this case, counterrun unrolling can help producing additional assumptions.

Counterrun unrolling consists in making the first traversals of looping states explicit. It is achieved by augmenting a counterplay with replicates of the looping states. The number of unrollings is referred to as the \emph{unrolling degree} $u$. Each unrolling  yields a new  set of states $\unrolledstates = \{\unrolledstate{1}{1},\dots,\unrolledstate{h}{1},\dots,$ $\unrolledstate{1}{u},$ $\dots,\unrolledstate{h}{u}\}$. An unrolled looping counterrun has the form $\initstate\transstate{1}\dots\transstate{k}$ $\unrolledstate{1}{1}\dots\unrolledstate{h}{1}\dots\unrolledstate{1}{u}\dots \unrolledstate{h}{u}(\loopingstate{1}\dots\loopingstate{k})^\omega$. 
Unrolling has two possible effects on the computed interpolant: on one hand, it can introduce new state components in the interpolant, which yield new invariant refinements according to (\ref{eq:Refinements}); on the other hand, the interpolant can express a more specific characterization of looping states, which corresponds to a weaker fairness refinement in (\ref{eq:Refinements}). These effects are both observed in our evaluation (see Sect.~\ref{sec:Evaluation}).

The translation from the temporal to the Boolean domain (lines \ref{lalg:unrTranslateCounterrunAssumptions}-\ref{lalg:unrTranslateGuarantees}) works as in Algorithm~\ref{alg:InterpolationBasedSynthesis}. When an interpolant is computed, provided that it is fully separable, it can contain components referring to the unrolled states. The state component referring to the $r$-th replica of the $j$-th looping state is denoted by $\boolexpr{unr}{j,r}{\var(\unrolledstate{j}{r})}$. The function \textbf{TranslateUnrolledInterpolant} (line \ref{lalg:unrTranslateInterpolant}) produces the formula
\begin{equation}
\label{eq:unrInterpolantTranslation}
\begin{split}
	\mathcal{T}(I_u) := & \boolxproj{init}{}{\inx} \land 
	\bigwedge_{s \in \states_r} \eventually \left( \boolexpr{}{s}{\var} \land \boolxproj{}{\successor{(s)}}{\next \inx} \right) \land \\
	&\eventually \always \bigvee_{j=1}^{|\loopingstates|} \left(\boolexpr{loop}{j}{\var} \land \bigwedge_{r=1}^u \boolexpr{unr}{j,r}{\var} \right)
\end{split}
\end{equation}
This formula is the same as \ref{eq:InterpolantTranslation} apart from the $\eventually\always$ conjunct. This is replaced with a stronger disjunction, where each disjunct groups the state components referring to all the replicates of the same looping state. By negating (\ref{eq:InterpolantTranslation}), the algorithm produces
\begin{equation}
\label{eq:unrRefinements}
\begin{split}
\lnot \boolxproj{init}{}{\inx} \lor 
\bigvee_{s \in S_u} \always \lnot \left( \boolexpr{}{s}{\var} \land \boolxproj{}{\successor{(s)}}{\next \inx} \right) \lor \\
\always \eventually \bigwedge_{j=1}^{|\loopingstates|} \lnot \left(\boolexpr{loop}{j}{\var} \land \bigwedge_{r=1}^u \boolexpr{unr}{j,r}{\var} \right)
\end{split}
\end{equation}
where each disjunct is a candidate refinement, and the fairness condition has a weaker form than the one in ($\ref{eq:Refinements}$).

Unrolling is stopped when some user-defined stopping condition is reached. This is checked by the function \textbf{CheckStoppingCondition}. Possible stopping conditions are:
\begin{itemize}
	\item the interpolant has yielded the same refinements as the previous step;
	\item no new refinements have been produced in the last $k$ unrolling step;
	\item a maximum unrolling degree has been specified.
\end{itemize}

%Following this, the process described  in  Sect.s~\ref{subsec:refcomp}--\ref{subsec:Unrolling} is repeated with new refinements computed. This continues  until weakest environment assumption (computable using our approach) that excludes the counterstrategy is  synthesised. The reason this results in a weaker assumption is that loop unrolling has the effect of delaying the occurrence of the failing loop.
%%, as shown in Figure~\ref{fig:UnrollingBound}. 
%The consequence is that  the computed interpolant may refer to additional state components compared to the one generated before the last unrolling occurred. This leads to generating additional candidate refinements at every unrolling step according to Formula (\ref{eq:Refinements}).

%%\begin{figure}
%%\centering
%%\includegraphics[width=0.4\linewidth]{figs/UnrollingBound}
%%\caption{Effect of loop unrolling on interpolants}
%%\label{fig:UnrollingBound}
%%\end{figure}

  %Notice that the iterative procedure of equivalence checking and computation from unrolled runs can viewed as an embedded oracle-guided synthesis cycle \cite{Seshia2015}, within an encompassing counterstrategy-guided refinement procedure.

\section{Convergence}
\label{sec:Convergence}
Our procedure is guaranteed to  terminate after a finite number of iterations. We  discuss below the case of all  computed interpolants being fully-separable. If at some step the interpolant is not fully separable, the \textbf{InterpolationBasedSynthesis} procedure returns the vacuous refinement $\false$, which ensures that branch of the refinement tree is not expanded further.

\begin{theorem}
Given a satisfiable but unrealizable specification $\langle  \phi^\env, \phi^\sys\rangle$ Algorithm~\ref{alg:Refinements}   terminates with a realizable specification $\langle  \phi^{\env'} , \phi^\sys\rangle$.
\end{theorem}

\proof To prove this, it is sufficient to show that the iteration in Algorithm~\ref{alg:Refinements} reaches its termination conditions. The argument we carry out mirrors the one for proving the termination of Abstract Learning Framework algorithms presented in \cite{Loding2016}.

In the following arguments, we will refer to the recursion tree of Algorithm~\ref{alg:Refinements}. Each node is associated with the candidate assumption tested in one specific call of \textbf{CounterstrategyGuidedRefinements}. The root corresponds to the initial assumption; every internal node symbolizes an unrealizable assumptions refinement; the children of an internal node correspond to the alternative refinements that rule out the relevant counterstrategy. The leaves represent  alternative realizable assumption refinements returned by the algorithm. We will show that this tree has finite depth and breadth.

%\subsection{Termination of the counterstrategy-guided recursion}
%\label{sec:TerminationCSGR}
%Let us first consider the former one. The number of recursive calls coincides with the number of nodes in the recursion tree in Figure~\ref{fig:recursiontree}. The internal nodes correspond to the computation of a counterstrategy and of the relevant refinements, while the leaves represent the last recursive call for a branch, where no computation is performed.

%The number of nodes in the tree is finite. 
Let us consider the number of children $n_\counterstrategy$ of an internal node (the subscript $\counterstrategy$ refers to the counterstrategy computed in that internal node). It corresponds to the maximum number of refinements that are generated from a single counterstrategy. Formula (\ref{eq:Refinements}) allows the extraction of one initial condition, one fairness condition and $|\states_p|$ invariants; therefore:
$$n_\counterstrategy = |\states_r| + 2 \;.$$

%Assuming that the maximum unrolling degree is finite (we will see that later in this section), denoted $u_{C,MAX}$, the maximum number of refinements generated from $C$ can be computed by counting the maximum number of disjuncts in (\ref{eq:Refinements}). Suppose $|S_{u_{C,MAX}}|$ denotes the number of distinct states in the unrolled counterplay, then $n_C \leq |S_{u_{C,MAX}}| + 2$: we count one initial condition, one fairness condition and $|S_{u_{C,MAX}}|$ invariants. Given that every node has a finite number of children, the breadth of each level in the tree is also finite.

We now consider the depth. The algorithm keeps refining a computed assumption until the property becomes realizable  (in case the returned refinement is \emph{false}, then the property is  realizable). 
%see Figures~\ref{fig:refinements}-\ref{fig:recursiontree}). 
Given the soundness property (Theorem~\ref{th:soundness}), at each step every refinement excludes the latest computed counterstrategy; since this counterstrategy satisfies all the previously computed refinements by definition, the new refinement cannot be equivalent to any of the previous refinements along the same branch. 
%For instance, in Figure~\ref{fig:recursiontree} $\psi_{2,1}$ cannot be equivalent to $\psi_2$ nor to any of the assumptions already in $\phi_e$.

For the above reason, the depth $d$ of the refinement tree is limited by the maximum number of existing GR(1) refinements modulo logical equivalence. The maximum number of initial conditions is 
$d_{init,MAX} = 2^{2^{|\mathcal{X}|}}$,
that is the number of all distinct Boolean expressions over the input variables. The maximum number of invariants is
$d_{inv,MAX} = 2^{2^{|\mathcal{V}|}+2^{|\mathcal{X}|}}$;
this corresponds to the maximum number of distinct $B_s$ that can be present in the expression (\ref{eq:Refinements}) times the number of distinct $B_{\successor{(s)},X}$. Finally, the maximum number of distinct fairness assumptions is
$d_{fair,MAX} = 2^{2^{|\mathcal{V}|}}$
Therefore, the total depth $d$ is bounded by the sum of these three quantities:
%\begin{equation}
%\label{eq:TreeDepth}
	$d \leq d_ {MAX} = d_{init,MAX} + d_{inv,MAX} + d_{fair,MAX}$. \qed
\endproof
%\end{equation}

Given the above, we conclude that the recursion tree is finite. This gives us a worst-case upper bound on the depth $d$ of the recursion, which has a doubly exponential growth over $|\mathcal{V}|$ --- a general observation of counterstrategy-guided assumptions refinement strategies. 
%
%In Chapter~\ref{ch:Evaluation} we show that our approach consistently eliminates an unrealizable core from the specification. If this observation can be proved to be a general property of refinements computed via our approach, it is possible to define a tighter upper bound on the tree depth. A necessary and sufficient condition to achieve realizability is that there are no subsets of minimally unfulfillable guarantees (see Section ~\ref{sec:Unrealizability}) after the refinement. So, the maximum number of refinements needed to achieve realizability under the above assumption is:
%\begin{equation}
%	d = N_G
%\end{equation}
%where $N_G$ is the number of subsets of minimally unfulfillable guarantees in the initial property.
%
%\subsection{Termination of the unrolling loop}
It remains to show that the inner-cycle terminates in finite time. As mentioned in Sect.~\ref{subsec:Unrolling}, each iteration can provide additional or weaker refinements with respect to the previous iteration. The termination condition holds when the current iteration does not yield new refinements with respect to the previous one. This  is reached in the worst case after all distinct GR(1) refinements are generated. The computation is the same as the one for $d$: $u_{C,MAX} = d_{MAX}$.
%\vspace{-4pt}
%\begin{equation}
%u_{C,MAX} = d
%\end{equation}
%\vspace{-4pt}
%\section{Theoretical Results}
%
%We extend the definition of helpfulness of an assumption with respect to a particular explanation.
%\begin{definition}
%	\label{def:helpful_assumption}
%	Let  $\langle A, G\rangle$ be an unrealizable specification. The assumption {\em a} is said to be helpful (resp.~unhelpful) w.r.t.~to an explanation    $\langle A', G'\rangle$, where  $A'\subseteq A$ and $G'\subseteq G$, if $\langle A'\cup\{${\em a}$\}, G'\rangle$ is realizable (resp.~unrealizable).
%\end{definition}
%
%
%\begin{theorem}
%	Let $\langle A, G\rangle$ be an unrealizable specification and $\langle A', G'\rangle$, where  $A'\subseteq A$ and $G'\subseteq G$, an explanation for the unrealizability. Let {\em a} be an assumption computed by the algorithm described in Section \ref{} from the explanation $\langle A', G'\rangle$ and counter-trace $\pi$. Then assumption {\em a} is helpful  w.r.t.~the explanation $\langle A', G'\rangle$. 
%\end{theorem}
%
%\begin{proof}[Sketch]
%	Assume $a=\neg \tran(I_k)$ is an unhelpful assumption w.r.t.~$\langle A', G'\rangle$. Since $\langle A', G'\rangle$ is unrealizable, then  $\langle A' \cup \{a\}, G'\rangle$ is also  unrealizable. For the latter specification, this implies that there exists no Mealy transducer $M$ with input \minx\ and output \mouty\ such that $M \models  A'\wedge a\rightarrow G'$. In other words for any $M$, there exists a run $\sigma$ of $M$ such that $ A'\wedge a\wedge \neg G'$ is satisfied.  
%\end{proof}
\section{Evaluation}
\label{sec:Evaluation}
We apply our approach to the two popular benchmarks presented in \cite{Alur2013,Bloem2012,Konighofer2009}: a lift controller and ARM's AMBA-AHB protocol. In addition we consider a selection of three case studies provided by \cite{Kuvent2017} for testing JVTSes (see Section~\ref{sec:Related}).
%All the steps described in Section~\ref{ch:Contribution} are executed on these case studies. 

%For each case study, we report the maximum depth and breadth of the recursion tree, and an interpretation of some interesting refinements that are computed. Details are available at  \cite{Davide}.%Full details are available \cite{Cavezza}.

\subsection{Case studies}
Due to its small size that allows a thorough description, we use the lift controller case study as a further demonstration of the kind of refinements generated via interpolation, as long as an example of the usefulness of unrolling. We describe it in its dedicated section below. We instead conduct an extensive experimental campaign on the other case studies to compare the efficiency of Alur's approach from \cite{Alur2013} (to which we refer as \emph{multivarbias}) and ours in finding realizable refinements.

The \emph{Advanced High-performance Bus} (\emph{AHB}) is part of the \emph{Advanced Microcontroller Bus Architecture} (\emph{AMBA}) specification. It is an open-source communication protocol for on-chip devices through a shared bus. Devices are divided into \emph{masters}, which initiate a communication, and \emph{slaves}, which respond to requests. Multiple masters can request the bus simultaneously, but only one at a time can communicate through it. Masters and slaves constitute the environment, while the system is the bus arbiter implementing the protocol.
%%%
The specification of the AHB protocol is a GR(1) description of the protocol developed by ARM and summarized in \cite{Bloem2012}. We consider specifications for two, four and eight masters (AMBA02, AMBA04, AMBA08 respectively) which are realizable. The original specification is realizable: in order to obtain an unrealizable case study on which we can evaluate our approach, we remove the assumption $\always\eventually \hready$ as done in \cite{Alur2013,Li2011a}.

We tested our approach on additional case studies coming from the work in \cite{Kuvent2017} on Justice Violation Transition Systems. The case studies include:
\begin{itemize}
	\item a robot sorting Lego pieces by color (\emph{ColorSort});
	\item a mobile robot of humanoid shape (\emph{Humanoid});
	\item a robot with self-balancing capabilities (\emph{Gyro}).
\end{itemize}

Table~\ref{tab:TestCases} provides a summary of  both case studies. The columns \textbf{In} and \textbf{Out} contain the number of input and output variables in the specification alphabet respectively; \textbf{A} and \textbf{G} contain the number of assumptions and guarantees respectively. %; \textbf{MaxPlay} contains the maximum number of states in a counterplay among all the counterplays used in the refinement process; \textbf{MaxUnr} reports the maximum unrolling degree reached in any step of the approach before reaching the termination condition; \textbf{TreeDepth} corresponds to the depth of the recursion tree; \textbf{MaxAltRef} is the maximum number of alternative refinements computed to rule out any single counterstrategy (it corresponds to the maximum number of children of an internal node in the recursion tree); \textbf{\#Ref} shows the total number of refinement sets computed  that make the property realizable.

\begin{table}
%\vspace{-10pt}
\caption{Summary of case studies}
\label{tab:TestCases}
\centering
\begin{tabular}{c|cccc}
Specification & In & Out & A & G \\ %& MaxPlay & MaxUnr & TreeDepth & MaxAltRef & \#Ref \\
\hline Lift & 3 & 3 & 7 & 12 \\ %& 2 & 2 & 1 & 3 & 3\\ 
 AMBA02 & 7 & 16 & 10 & 66 \\ %& 4 & 2 & 3 & 6 & 17\\ 
 AMBA04 & 11 & 23 & 16 & 97 \\ %& 7 & 1 & 2 & 2 & 8\\ 
 AMBA08 & 19 & 36 & 28 & 157 \\ %& 18 & 1 & 7 & 2 & 80\\
 ColorSort & 18 & 12 & 10 & 21 \\
 Humanoid & 6 & 16 & 0 & 23 \\
 Gyro & 6 & 4 & 10 & 8 
\end{tabular}
%\vspace{-6pt}
\end{table}
%The objective is to provide examples of application of our approach, and to highlight the properties discussed in Section~\ref{ch:Contribution}.
%
%In this section we will provide the main observations on the output. The full execution of the procedure on the examples is given in the appendix.

\subsection{Lift Controller}
This case study (also used for controller synthesis problems \cite{Bloem2012,Alur2013}) involves the specification of a system comprising a lift controller. The lift moves between three floors. The environment consists of three buttons, whose states can be \emph{pressed} or \emph{unpressed}; the corresponding state is represented by three binary input variables $\{b_1,b_2,b_3\}$. The controller's state consists of three output variables $\{f_1, f_2,f_3\}$ that indicate at which floor the lift is.
The assumptions are:
\begin{enumerate}
	\item $\varphi^e_{init} = \lnot b_1 \wedge \lnot b_2 \wedge \lnot b_3$
	\item $\varphi^e_{1,i} = \always (b_i \wedge f_i \rightarrow \next \lnot b_i)$
	\item $\varphi^e_{2,i} = \always(b_i \wedge \lnot f_i \rightarrow \next b_i)$
\end{enumerate}
for $i \in \{1,2,3\}$.
They state that the buttons are not pressed in the initial state (1);  a pressed button transits to a non-pressed state when the lift arrives at the corresponding floor (2); and the button remains in the pressed state until the lift arrives at that floor (3).
The guarantees are:
\begin{enumerate}
	\item $\varphi^s_{init} = f_1 \wedge \lnot f_2 \wedge \lnot f_3$
	\item $\varphi^s_1 = \always (\neg (f_1 \wedge f_2) \wedge \neg (f_2 \wedge f_3) \wedge \neg (f_1 \wedge f_3))$
	\item $\varphi^s_{2,1} = \always (f_1 \rightarrow (\next f_1 \vee \next f_2))$
	\item $\varphi^s_{2,2} = \always (f_2 \rightarrow (\next f_1 \vee \next f_2 \vee \next f_3))$
	\item $\varphi^s_{2,3} = \always (f_3 \rightarrow (\next f_2 \vee \next f_3))$
	\item $\varphi^s_3 = \always (((f_1 \wedge \next f_2) \vee (f_2 \wedge \next f_3) \vee (f_2 \wedge \next f_1) \vee (f_3 \wedge \next f_2)) \rightarrow (b_1 \vee b_2 \vee b_3))$
	\item $\varphi^s_{4,i} = \always\eventually(b_i \rightarrow f_i)$
	\item $\varphi^s_{5,i} = \always\eventually f_i$
\end{enumerate}
for $i \in \{1,2,3\}$.
They state that the lift starts from floor $1$ (1);  it can never be in two floors at the same time (2);  it can move only between consecutive states (3-5), and moves only when at least a button is pressed (6);  plays in which the environment keeps a button $b_i$ pressed infinitely and the lift never reaches the corresponding $f_i$ are forbidden (7); and that the lift is required to visit all the floors infinitely often (8). Given this specification, the fairness guarantee can be satisfied if the environment sets one of its $b_i$ to 1 at least once.

\begin{figure}
%\vspace{-10pt}
\centering
\includegraphics[width=.4\textwidth]{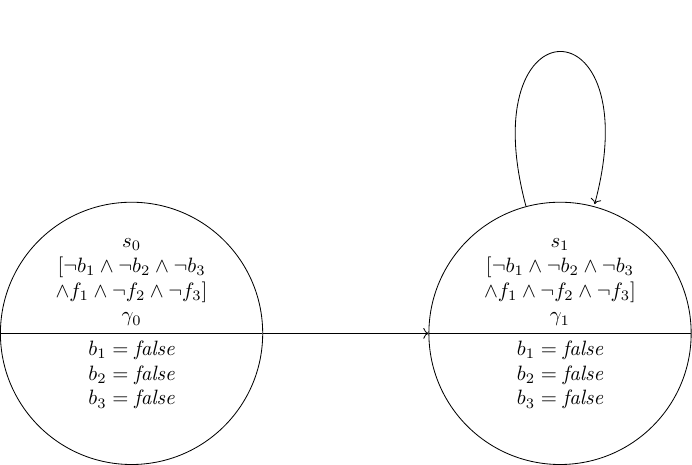}
%\vspace{-6pt}
\caption{Lift counterstrategy produced by RATSY}
\label{fig:graphlift}
\vspace{-10pt}
\end{figure}

The specification is unrealizable, since when the buttons (environment) stay indefinitely unpressed, the lift (controller) cannot move and therefore $\varphi^s_{5,2}$ and $\varphi^s_{5,3}$ are violated. The unrealizable core  consists of the whole set of assumptions and the guarantees $\varphi^s_{init}$, $\varphi^s_{2,1}$, $\varphi^s_{3}$ and $\varphi^s_{5,2}$. From this core, RATSY computes the counterstrategy $\counterstrategy$ in Fig.~\ref{fig:graphlift}, which consists of a unique run $r_\counterstrategy$. After translating the unrealizable core over the counterplay, the interpolant is $I_0 = \lnot b_1(s_0) \land \lnot b_2(s_0) \land \lnot b_3(s_0),$ which corresponds to the GR(1) refinements $\lnot b_1 \land \lnot b_2 \land \lnot b_3$. However, this assumption is inconsistent with $\varphi^e_{init}$, and therefore makes the specification vacuously realizable.

\begin{figure}
	%\vspace{-10pt}
	\centering
	\includegraphics[width=.7\textwidth]{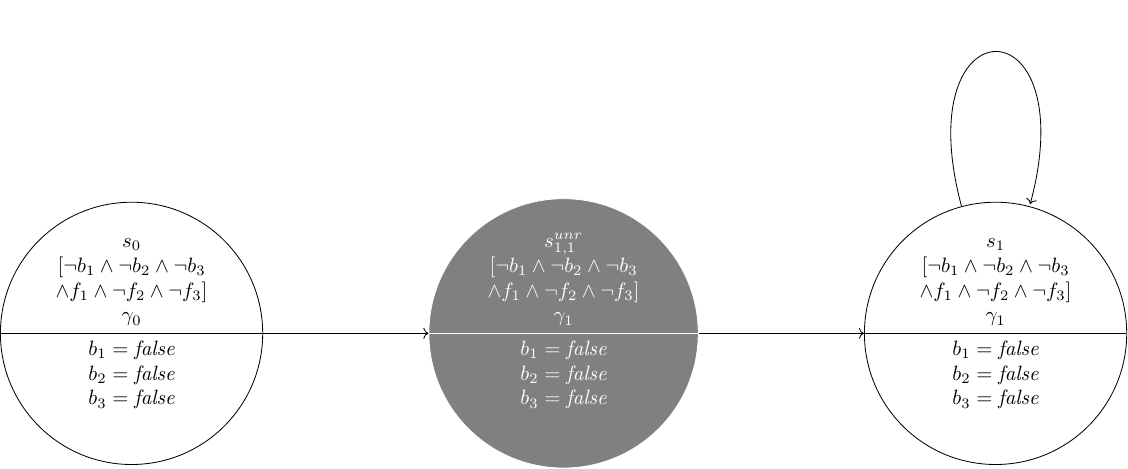}
	%\vspace{-6pt}
	\caption{Unrolled counterrun}
	\label{fig:unrolledgraphlift}
	\vspace{-10pt}
\end{figure}

We resort to unrolling for obtaining more alternative refinements. The resulting counterrun is shown in Fig.~\ref{fig:unrolledgraphlift}. After unrolling, the procedure yields the interpolant $I_1 = \lnot b_1(s_0) \land \lnot b_2(s_0) \land \lnot b_3(s_0) \land \lnot b_1(s^{unr}_{1,1}) \land \lnot b_2(s^{unr}_{1,1}) \land \lnot b_3(s^{unr}_{1,1})$.
%, which is translated as $$\lnot b_1 \land \lnot b_2 \land \lnot b_3 \land \mathbf{F}(\lnot b_1 \land \lnot b_2 \land \lnot b_3 \land \next(\lnot b_1 \land \lnot b_2 \land \lnot b_3)) \land \mathbf{FG}(\lnot b_1 \land \lnot b_2 \land \lnot b_3)).$$
By translating and negating this interpolant, we obtain the refinements
\begin{enumerate}
	\item $b_1 \lor b_2 \lor b_3$
	\item $\always(\lnot b_1 \land \lnot b_2 \land \lnot b_3 \rightarrow \next(b_1 \lor b_2 \lor b_3))$
	\item $\always\eventually(b_1 \lor b_2 \lor b_3)$
\end{enumerate}
Notice that unrolling results in an interpolant containing an additional state component, thus allowing for more alternative refinements (see Sect.~\ref{subsec:Unrolling}). Moreover, the new state component refers to an unrolled state, from which 
a new fairness refinement not inferable from $I_0$ is generated.
%
%The second unrolling produces equivalent refinements, and thereby the inner-cycle terminates.  %Notice that the observation about unrolling stated in Section~\ref{sec:Complexity} is confirmed here: the second unrolling includes the description of $s^{unr}_{1,2}$, which is the second replica of $s_1$; since a replica of $s_1$ was already present in the interpolant $I_1$, the new unrolling does not generate any new refinement.

Every candidate refinement computed by our approach eliminates an unrealizable core.
Moreover, each one solves the unrealizability problem for the original specification. Refinement (1) does this in a trivial way, since it contradicts the initial assumption contained in the specification.
Notice that all the computed refinements force at least one of the buttons to be pressed at some point in any play of the environment. This %is consistent with the solution to unrealizability we envisioned, and 
corresponds to the refinement output by the approach in \cite{Alur2013}.
%

%\begin{table}
%	\centering
%	\caption{Lift controller refinement}
%	\label{tab:lift}
%	\begin{tabular}{|c|c|}
%		\hline
%		Nr. input variables & 3 \\
%		Nr. output variables & 3 \\
%		Nr. assumptions & 7 \\
%		Nr. guarantees & 12 \\
%		%Nr. guarantees (unreal. core) & 4 \\
%		Max path length & 2 \\
%		Max unrolling degree & 2 \\
%		Recursion tree depth & 1 \\
%		Recursion tree breadth & 3 \\
%		Refinements & 3 \\\hline
%	\end{tabular}	
%\end{table}

\subsection{Experimental setup}
To compare our interpolation-based approach with the one in \cite{Alur2013}, we run both on the case studies in Table~\ref{tab:TestCases}. The two approaches are equal in the search strategy of the refinement tree, that is a breadth-first search (see Algorithm~\ref{alg:Refinements}). They differ in the way refinements are generated from counterstrategies. Since the latter's generation depends on the choice of the variables to use in the refinements, and we assume no a priori knowledge about which variables need to be chosen to achieve realizability, we use five distinct subsets randomly selected at each refinement step. We call this approach \emph{multivarbias}, as the search is biased by multiple variable choices.

Each of the approaches is run on each case study for up to 24 hours, to allow sufficient time for an analysis of the trend in generating realizable refinements. The requirements analysis tool RATSY \cite{Bloem2010} is used to check unrealizability and compute counterstrategies. The SAT solver MathSAT \cite{Cimatti2008,Bruttomesso2008} is used to compute interpolants. The refinement synthesis procedure is implemented in Python 2.7. The experiment is executed on a dedicated Ubuntu machine with an Intel Core i7 CPU and 16 GiB of memory.

During the search we record the time at which every refinement node is generated, whether or not the refinement makes the specification realizable, whether it eliminates the parent's unrealizable core, and whether the refinement is satisfiable or is equivalent to the constant $\false$ (see Section~\ref{subsec:Unrolling} about vacuously realizable specifications). To our knowledge, this is the first comparative assessment of refinement approaches based on a quantitative analysis of performance over execution time.

\subsection{Results}
\begin{figure}[t!]
	\centering
	\begin{subfloat}{
		\centering
		\includegraphics[width=0.46\linewidth]{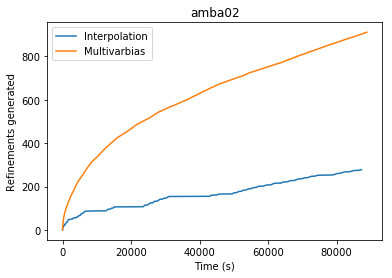}
	}
	\end{subfloat}
	\begin{subfloat}{
			\centering
			\includegraphics[width=0.46\linewidth]{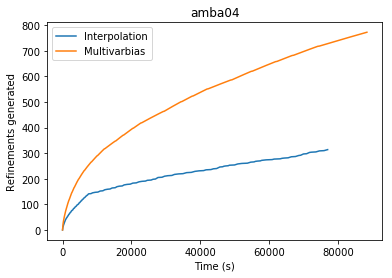}
		}
	\end{subfloat}\\
	\begin{subfloat}
		\centering
		\includegraphics[width=0.46\linewidth]{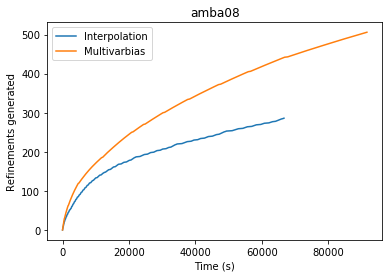}
	\end{subfloat}
	\begin{subfloat}
		\centering
		\includegraphics[width=0.46\linewidth]{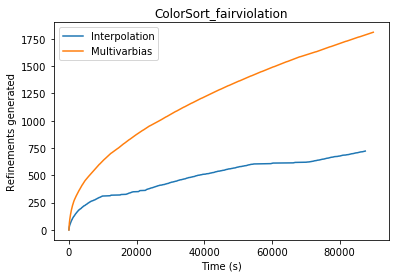}
	\end{subfloat}\\
	\begin{subfloat}
		\centering
		\includegraphics[width=0.46\linewidth]{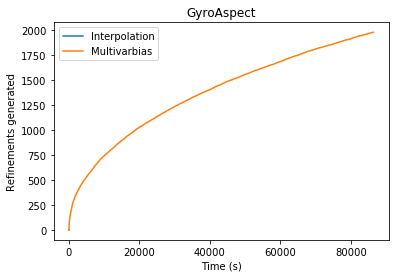}
	\end{subfloat}
	\begin{subfloat}
		\centering
		\includegraphics[width=0.46\linewidth]{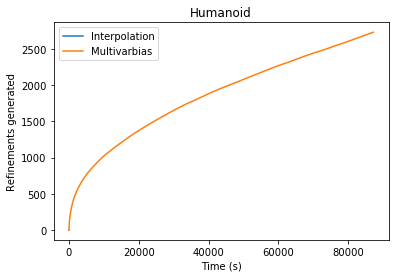}
	\end{subfloat}
	\caption{Total number of refinement nodes explored over time}
	\label{fig:TotalRefs}
\end{figure}

\begin{figure}[t!]
	\centering
	\begin{subfloat}{
			\centering
			\includegraphics[width=0.46\linewidth]{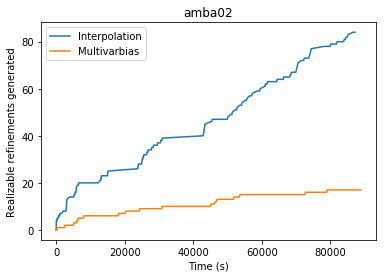}
		}
	\end{subfloat}
	\begin{subfloat}{
			\centering
			\includegraphics[width=0.46\linewidth]{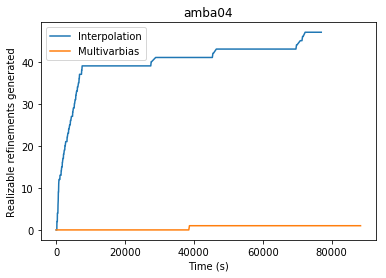}
		}
	\end{subfloat}\\
	\begin{subfloat}
		\centering
		\includegraphics[width=0.46\linewidth]{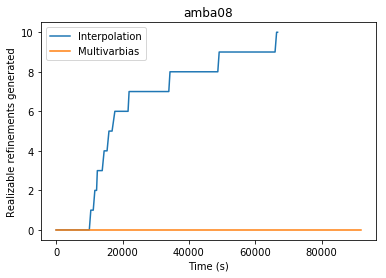}
	\end{subfloat}
	\begin{subfloat}
		\centering
		\includegraphics[width=0.46\linewidth]{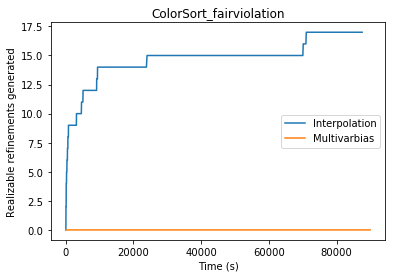}
	\end{subfloat}\\
	\begin{subfloat}
		\centering
		\includegraphics[width=0.46\linewidth]{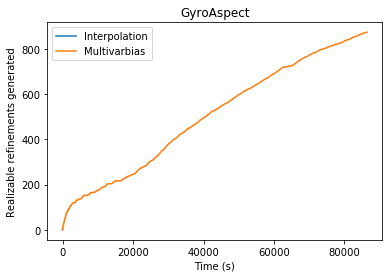}
	\end{subfloat}
	\begin{subfloat}
		\centering
		\includegraphics[width=0.46\linewidth]{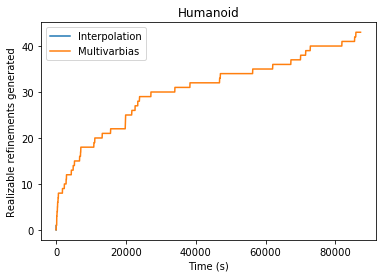}
	\end{subfloat}
	\caption{Number of realizable refinement nodes discovered over time (excluding vacuously realizable ones)}
	\label{fig:RealizableRefs}
\end{figure}

\begin{figure}[t!]
	\centering
	\begin{subfloat}{
			\centering
			\includegraphics[width=0.46\linewidth]{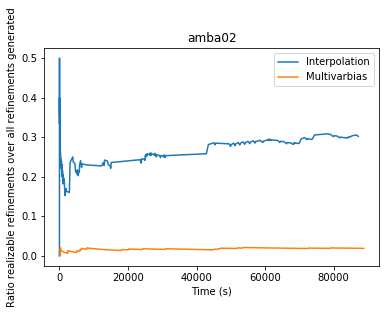}
		}
	\end{subfloat}
	\begin{subfloat}{
			\centering
			\includegraphics[width=0.46\linewidth]{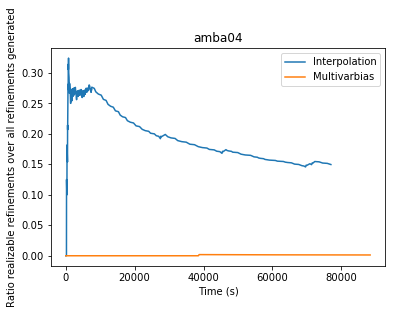}
		}
	\end{subfloat}\\
	\begin{subfloat}
		\centering
		\includegraphics[width=0.46\linewidth]{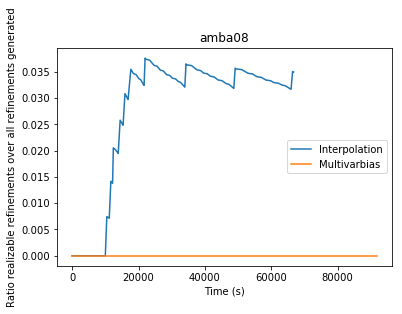}
	\end{subfloat}
	\begin{subfloat}
		\centering
		\includegraphics[width=0.46\linewidth]{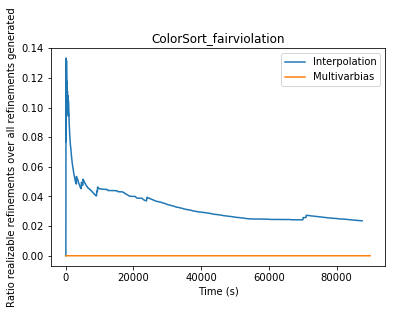}
	\end{subfloat}\\
	\begin{subfloat}
		\centering
		\includegraphics[width=0.46\linewidth]{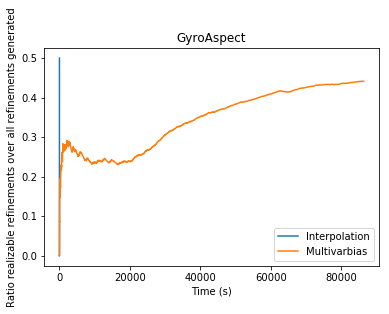}
	\end{subfloat}
	\begin{subfloat}
		\centering
		\includegraphics[width=0.46\linewidth]{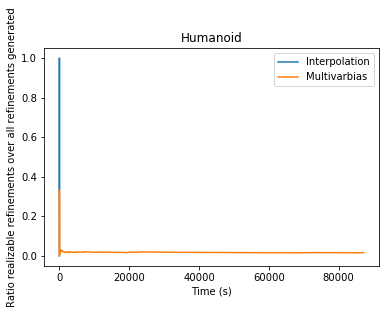}
	\end{subfloat}
	\caption{Ratio between realizable nodes and total generated nodes over time}
	\label{fig:RatioRefs}
\end{figure}

Figures~\ref{fig:TotalRefs}-\ref{fig:RatioRefs} summarize the results of the experiment on each of the six case studies. Fig.~\ref{fig:TotalRefs} shows the total number of refinement tree nodes searched by both approaches. Notice that in all case studies a larger number of nodes is explored by multivarbias than with interpolation; this is due to the additional overhead needed to compute interpolants with respect to template instantiation \cite{Alur2013}. We also point out that on the GyroAspect and Humanoid case studies, the interpolation-based approach terminates after less than 10 seconds, while multivarbias continues for the whole time frame of the experiment.

Fig.~\ref{fig:RealizableRefs} shows the number of realizable refinements discovered over time.  In all case studies, the curve of interpolation remains above the one of multivarbias for the entire time frame where both approaches are executing. Therefore, it appears evident that interpolation is more focused on realizable nodes than multivarbias. This observation is corroborated by Fig.~\ref{fig:RatioRefs}, which shows the proportion of realizable refinements discovered over all explored nodes over time. The interpolation approach settles around a value bigger than 15\% in the first four case studies, while multivarbias remains close to 0\%. In addition, the latter does not find any realizable refinement at all in the AMBA08 and ColorSort case studies.

On GyroAspect, interpolation explores 4 nodes, out of which it finds 2 solutions; on Humanoid, 3 nodes are explored, all of them being solutions to realizability; the total exploration time is less than 10 seconds. On the other hand, multivarbias finds 873 solutions on GyroAspect and 43 solutions on Humanoid by running throughout the entire 24-hour time frame. In both cases, the maximum ratio of realizable solutions over the total explored nodes is achieved by interpolation.

\subsection{Discussion}
A reason for the efficiency of interpolation can be sought in its addressing unrealizable cores directly. Given a GR(1) specification $\langle \phi^\env,\phi^\sys \rangle$ with unrealizable core $\unrealcore$, and a refinement $\psi$, we say that $\psi$ \emph{targets} $\unrealcore$ if and only if the specification $\langle \phi^\env \cup \{\psi\}, \unrealcore$ is realizable. Intuitively, any unrealizable core must be targeted by a refinement that solves the unrealizability problem. Therefore, when building the refinement tree it is desirable that at least an unrealizable core be targeted by a partial refinement.

Table~\ref{tab:TargetingUnrealCores} shows the percentage of nodes targeting their respective parent's unrealizable core violated by the produced counterstrategy (see Section~\ref{sec:Background} for the relationship between counterstrategies and unrealizable cores). By comparing this with Fig.~\ref{fig:RatioRefs}, notice that for each approach and each case study the procedure is most efficient in identifying realizable refinements if the proportion of nodes targeting their parent's unrealizable core is higher on that case study.
\begin{table}
	\caption{Percentage of nodes targeting parent's unrealizable cores}
	\label{tab:TargetingUnrealCores}
	\centering
	\begin{tabular}{c|cc}
		& Interpolation & Multivarbias \\ 
		\hline 
		AMBA02 & 88.17\% & 5.48\% \\ 
		AMBA04 & 100\% & 1.03\% \\ 
		AMBA08 & 100\% & 0.39\% \\ 
		ColorSort & 45.09\% & 1.6\% \\ 
		GyroAspect & 100\% & 80.5\% \\ 
		Humanoid & 100\% & 100\% \\ 
	\end{tabular}
\end{table}
Therefore, producing refinements from interpolants (which in turn are computed from the translation of an unrealizable core), amounts to performing an educated choice of the variables to use in the refinements to be generated.

The larger number of solutions found by multivarbias in GyroAspect and Humanoid may be related to the size of the initial specification: the last two case studies are the smallest in terms of number of variables; therefore the space of possible choices for variables is small, and the likelihood of choosing a set of variables leading to a solution is higher. However, the number of variables alone may not be sufficient to characterize a class of problems where random variable selection works better than interpolation; in fact, AMBA02 has the same size in terms of number of variables, but in that case interpolation shows better performance than multivarbias.

The other difference to be considered lies in the number of initial assumptions and guarantees, which is higher in AMBA02 than in the other two case studies. This leads to longer translations from LTL to Boolean, which in turn potentially lead to longer interpolants and therefore more nodes in interpolation's refinement tree. Therefore, in characterizing such class of problems the number of initial assumptions and guarantees plays a central role as well. Further studies are to be conducted in this direction.

\section{Discussion}
\label{sec:Discussion}
A direction of improvement of our approach lies in the selection of the counterrun to generate refinements from. Currently this choice is made at random, and only one counterrun is selected at every refinement synthesis step. Alternative approaches can be devised where more than one counterrun is selected, and/or computable heuristics of counterruns are exploited in order to focus on the ones with the largest number of alternative solutions. The use of heuristics would mitigate the effect of generating more intermediate nodes in the refinement tree that do not lead to realizable specifications.

Recently Justice Violations Transition Systems (JVTS) \cite{Kuvent2017} have been proposed as alternatives to counterstrategies for debugging unrealizable specifications. The authors show that these structures are more concise and their computation is more time-efficient than counterstrategies, while retaining the same expressive power. However, to our knowledge no refinement approach has been proposed to date that uses such structures. In our work, we focus on counterstrategies in order to provide a fairer comparison with the state of the art. Investigation on how interpolation can be integrated with JVTSes is matter for future work.
The use of JVTSes, however, by speeding up the counterstrategy generation phase, will allow to generate bigger refinement trees in a given time, possibly discovering a higher number of solutions.

%[Davide: why do you think this is relevant?]
%Our focus has been in GR(1) as the specification language for its relevance in Reactive synthesis from GR(1) for its common use, e.g., see \cite{Braberman2013a,DIppolito2014,Kress-Gazit2009}. More recent research has focused on a different fragment, called Mode Target \cite{Balkan2017}, whose synthesis is also polynomial in the length of the specification. [Davide: perhaps say here what would need to change to apply it to this fragment instead]
\section{Related Work}
\label{sec:Related}

\subsection{Revision of systems specifications}
GR(1) assumptions refinement is an instance of the general problem of LTL specification revision. This has applications in many contexts, like robot motion planning \cite{Fainekos2011,Kim2015},  operational requirements elaboration  \cite{Alrajeh2013a}.

 The work in \cite{DIppolito2014} takes a complementary perspective on GR(1) assumptions refinement by proposing a framework that weakens guarantees in order to adapt the functional behavior of a controller in accordance with observed violations of some given assumptions.

Other related work on assumption refinement includes those operating directly on  game structures  \cite{Chatterjee2008}. With regard to the parity game model used for controller synthesis (such as in \cite{Sohail2013,Sohail2008}), this paper defines the concept of safety assumptions as sets of edges that have to be avoided by the environment, and the concept of fairness assumptions as sets of edges that have to be traversed by the environment infinitely often. The work devises an algorithm for finding minimal edge sets in order to ensure that the controller has a winning strategy. Our approach instead focuses on synthesizing general declarative temporal assertions whose inclusion has the effect of removing  edges from the game structure.

The problem of synthesizing environment constraints  has  been tackled in the context of assume-guarantee reasoning for compositional model checking \cite{Pasareanu1999,Cobleigh2003a,GheorghiuBobaru2008} to support compositional verification. 
%Here the focus is in verification of systems rather than synthesis. 
In these,  assumptions are typically expressed as labeled transition systems (LTSs) and learning algorithms like ${\mathcal L}^*$ \cite{Angluin1987} are used to incrementally refine the environment assumptions needed in order to verify the satisfaction  of properties.

Add discussion here about abstraction refinement
 
Add discussion here Specification mining from execution traces

\subsection{Oracle-Guided Inductive Synthesis}
Countstrategy-guided assumptions refinement is an instance of the Oracle-Guided Inductive Synthesis (OGIS) framework \cite{Seshia2015,Jha2017}. This framework consists of an iterative interaction between a learner, which tries to learn some solution concept by induction over a set of examples, and an oracle, which answers the learner's queries by providing meaningful examples for the learning process. In our context, the solution concept to learn is a refinement that makes a GR(1) specification realizable; the learner is the refinement synthesis procedure (Algorithm~\ref{alg:InterpolationBasedSynthesis} described in Section~\ref{sec:Interpolation}); and the oracle is the realizability checking engine (invoked in line \ref{lalg:RealizabilityCheck} of Algorithm~\ref{alg:Refinements}).
%In this framework the solution to a synthesis problem is sought through submitting a sequence of hypotheses to an oracle, which responds with a counterstrategy in case the hypothesis does not match the solution.
%; the counterexample serves to drive the synthesizer to formulate a new hypothesis in a loop, which terminates when the solution has been reached or no hypotheses can be formulated. 
Other instances of OGIS are counterexample-guided inductive synthesis of programs \cite{Jha2010,Alur,Jha2014a} of abstraction refinement \cite{Clarke2000a,Esparza2006} and obstacle detection \cite{Alrajeh:2012}.

\subsection{Applications of Craig interpolation}
Craig interpolants have been deployed in the context of abstraction refinement for verification  in \cite{Esparza2006,DSilva2008}. The differences with our work are in specification language and overall objective: they seek additional assertions for static analysis of programs, while we look for GR(1) refinements of systems specifications to enable their automated synthesis. The authors of  \cite{Degiovanni2014a} use interpolation to support the extraction of pre- and trigger-conditions of operations within event-driven systems to enable the `satisfaction' of goals expressed within restricted  fragment of LTL. Though different in objective, approach and class of properties,  our technique can help in identifying specifications operationalizable by \cite{Degiovanni2014a}. This work also inspires our translation procedure from temporal logic to pure Boolean and vice versa.

%Finally, we validate the strengths of our approach using two benchmarks from the literature.
%Our contributions are summarised as follows. We propose a new  algorithm for synthesising environment assumptions  for GR(1) specifications based on Craig's interpolation. Our approach  directly targets counter-strategies and unrealizable cores, and does not require users to provide  variables for constructing these refinements. 
%
%We show, through its application, that our approach potentially converges more quickly to a realizable specification. %even if restricted to the variables  that lie in the shared language of the assumptions and guarantees. 
%We show, through its application, that targeting unrealizable cores prevents our procedure from generating refinements that do not help to get closer to realizability.
\section{Conclusions}
\label{sec:Conclusions}

We presented an interpolation-based approach for synthesizing weak environment assumptions for GR(1) specifications. 
Our approach exploits the information in counterstrategies and unrealizable cores to compute
assumptions that  directly target the cause of unrealizability.  Compared to closely related approaches \cite{Alur2013,Li2011a}, our algorithm does not
require the user to provide the set of variables upon which the assumptions are constructed. The case study applications show that our approach implicitly performs a variable selection that targets an unrealizable core, allowing for a quicker convergence to a realizable specification.
%converges more quickly compared to these, even when restricted to the shared alphabet $\lang_{\phi^\env} \cap \lang_{\phi^\sys} $. To the best of our knowledge, the use of interpolation for counterstrategy-guided synthesis and the exploitation of unrealizable core information have not been done before.

The final set of refinements is influenced by the choice of counterplay. We are investigating in our current work the effect of and criteria over the counterplay selection particularly on the full-separability of interpolants.  
Furthermore, since interpolants are over-approximations of the counterplays, the final specification is an under-approximation. In future work, we will explore the use of witnesses (winning strategies for the system) to counteract this effect. Finally, the applicability of our approach depends on the separability properties of the computed interpolants: further investigation is needed to characterize the conditions under which an interpolation algorithm returns fully-separable interpolants.
%We have  started to develop a fully integrated tool for our synthesis approach which we aim to make publicly available. 

\section*{Acknowledgments}
The support of the EPSRC HiPEDS CDT (EP/L016796/1) and Imperial College Junior Research Fellowship is gratefully acknowledged.

\newpage
\bibliographystyle{spbasic}
\bibliography{MyCollection}

%\newpage
%
%\input{sections-v3/appendix}

\end{document}